\documentclass[aps, prd, reprint, amsfonts, amssymb, amsmath, groupedaddress, showkeys, floatfix]{revtex4-1}

\usepackage{hyperref} 
\usepackage{graphicx} 
\usepackage{subfig} 
\usepackage{hyphenat} 
\usepackage{xspace} 
\usepackage{MnSymbol} 
\usepackage{array} 
\usepackage{multirow} 

\newcommand{\xmax}{$X_{\text{max}}$\xspace}

\begin{document}

\title{Studying the mass sensitivity of air-shower observables using simulated cosmic rays}
\author{Benjamin Flaggs}
\email{bflaggs@udel.edu}
\author{Alan Coleman}
\email{alanc@udel.edu}
\altaffiliation[now at]{ Department of Physics and Astronomy, Uppsala University, Uppsala SE-752 37, Sweden.}
\author{Frank G.~Schr\"oder}
\email{fgs@udel.edu}
\altaffiliation[also at]{ Institute for Astroparticle Physics, Karlsruhe Institute of Technology, D-76021 Karlsruhe, Germany.}
\affiliation{Bartol Research Institute, Department of Physics and Astronomy, University of Delaware, Newark, DE 19716, USA.}

\date{\today}

\begin{abstract}
    Using CORSIKA simulations, we investigate the mass sensitivity of cosmic-ray air-shower observables for sites at the South Pole and Malarg\"ue, Argentina, the respective locations of the IceCube Neutrino Observatory and the Pierre Auger Observatory.
    Exact knowledge of observables from air-shower simulations was used to study the event-by-event mass separation between proton, helium, oxygen, and iron primary cosmic rays with a Fisher linear discriminant analysis.
    Dependencies on the observation site as well as the energy and zenith angle of the primary particle were studied in the ranges from $10^{16.0}-10^{18.5}\,$eV and $0^\circ$ to $60^\circ$: they are mostly weak and do not change the qualitative results.
    Promising proton-iron mass separation is achieved using combined knowledge of all studied observables, also when typical reconstruction uncertainties are accounted for.
    However, even with exact measurements, event-by-event separation of intermediate-mass nuclei is challenging and better methods than the Fisher discriminant and/or the inclusion of additional observables will be needed.
    As an individual observable, high-energy muons ($>$ 500\,GeV) provide the best event-by-event mass discrimination, but the combination of muons of any energy and \xmax provides already a high event-by-event separation between proton-iron primaries at both sites.
    We also confirm that the asymmetry and width parameters of the air-shower longitudinal profile, $R$ and $L$, are mass sensitive.
    Only $R$ seems to be suitable for event-by-event mass separation, but $L$ can potentially be used to statistically determine the proton-helium ratio.
    Overall, our results motivate the coincident measurement of several air-shower observables, including at least \xmax and the sizes of the muonic and electromagnetic shower components, for the next generation of air-shower experiments.
\end{abstract}


\maketitle


\section{Introduction} \label{sec:intro}

One of the most prominent questions in the field of cosmic-ray physics currently pertains to the origin and acceleration mechanisms of the highest energy cosmic rays~\cite{Coleman:2022abf,Sarazin:2019fjz, Schroder:2019rxp}.
Studying the cosmic-ray flux as it varies with energy, especially the mass composition of this flux, is imperative to fully understand these unknowns.
In addition, the mass measurement of individual cosmic rays is desirable as a presumed mass dependence of the weak cosmic-ray anisotropy may reveal information about the sources and the propagation.
At energies above a few $100\,$TeV, the cosmic-ray flux falls below the detectable threshold by high-altitude balloon and satellite direct detection methods.
Hence, the highest energy cosmic rays must be detected indirectly via extensive air showers of secondary particles and electromagnetic radiation.
Once an air shower is initiated, direct mass measurements of the primary cosmic ray are no longer possible; however, specific air-shower observables are statistically related to the mass of the primary cosmic ray.

The slant depth within the atmosphere at which the number of electromagnetic shower particles ($\text{e}^{+}$/$\text{e}^{-}$/$\gamma$) is maximum, referred to as the depth of shower maximum (\xmax), and a probe of the total muonic component of the shower ($N_{\mu}$) are both mass sensitive shower observables for a given air-shower energy.
Specifically, $\langle X_{\text{max}} \rangle \propto -\ln{(A)}$ and $\langle N_{\mu} \rangle \propto A^{\simeq 0.15}$ where $A$ is the mass number of the primary nucleus~\cite{Matthews:2005sd}.
Differences in atmospheric depth between \xmax and ground results in a direct relation between the depth of shower maximum and the relative electromagnetic and muonic particle numbers at ground.
Taking the ratio of the electromagnetic and muonic particle numbers at ground forms the electron-muon ratio of the air shower, which is known to be sensitive to the mass of the primary particle~\cite{Apel:2008hi}.
The shape of the air-shower profile is also known to be mass sensitive due to its dependence on the interaction cross section of the primary particle~\cite{Andringa:2011zz}.

These observables are only statistically related to the mass of the primary cosmic ray due to shower-to-shower fluctuations.
Hence, on average there is a correlation between these observables and cosmic-ray mass, although mass determination of single events is challenging due to intrinsic mass-separation limits for the individual observables.
This challenge can be alleviated by combining knowledge of multiple mass sensitive observables, which yields the best prospects for cosmic-ray mass determination.
This simulation study serves to investigate the intrinsic potential of cosmic-ray mass separation based on combined knowledge of these observables. An earlier study had a similar goal~\cite{Holt:2019fnj}, although investigated fewer parameters and a smaller energy range.

CORSIKA~\cite{Heck:1998vt} simulations are used to study cosmic-ray air-shower observables.
From these simulations, exact knowledge of the mass sensitive observables \xmax, $N_{\mu}$, electron-muon ratio, and shape of the shower profile are obtained.
Within our analysis, all $N_{\mu}$ observables will reference the air-shower muonic component at the height of a detector array on the ground as this is related to the total muonic component of the shower.
Using knowledge of these observables, the separation power between proton (fully ionized hydrogen), helium, oxygen, and iron cosmic rays is studied, along with the separation power dependencies on both cosmic-ray energy and zenith angle.
Although the purpose of this simulation study is to show the intrinsic, general potential of different shower observables for event-by-event mass separation, CORSIKA simulations of air showers necessarily need to be done for specific conditions, such as the observation altitude, the atmospheric properties, or local geomagnetic field. 

Motivated by ongoing and planned upgrades of the IceCube Neutrino Observatory~\cite{IceCube:2016zyt} and the Pierre Auger Observatory~\cite{PierreAuger:2015eyc}, we have chosen their locations for this simulation study.
For comparability, the chosen energy range in both cases is the same from $10^{16.0}-10^{18.5}\,$eV, which is the energy range of the transition from the most energetic Galactic cosmic rays to extragalactic cosmic rays.
IceCube-Gen2~\cite{IceCube-Gen2:2020qha} plans to deploy a surface array of radio antennas and scintillator panels over an area of approximately 6\,$\text{km}^{2}$, overlapping with the current IceTop~\cite{IceCube:2012nn} footprint, allowing for mass sensitive observable measurements from air showers within the studied energy range~\cite{Haungs:2019ylq, IceCube-Gen2:2021jce}.
AugerPrime reaches to higher energies, but also covers the energy range from a few $10^{16.0}\,$eV upwards with the underground muon detectors and the 433\,m infill array of AMIGA~\cite{PierreAuger:2020gxz,PierreAuger:2021fhj}.
Currently being deployed as part of AugerPrime are a scintillator panel and radio antenna at the site of each water-Cherenkov surface detector within the existing 3000\,$\text{km}^{2}$ array~\cite{PierreAuger:2016qzd}, which will provide improved measurements of mass sensitive air-shower observables.
As expected from the universality of air showers, the results of both locations are overall consistent, which suggests that the simulation study is indeed of general value and can be used for the planning of future air-shower arrays with yet open locations.


\section{Cosmic-Ray Air-Shower Simulations} \label{sec:simulations}

All cosmic-ray air-shower simulations used in this analysis were generated as general purpose simulation libraries, with uses beyond this mass sensitivity analysis. 

CORSIKA v7.7401 was used to simulate the cosmic-ray air showers at the site of the IceCube Neutrino Observatory by setting the observation level of the simulations to 2840\,m above sea level, the magnetic field to 16.75\,\textmu T in the geomagnetic North direction and $-$51.96\,\textmu T downwards, and the atmosphere to model 33 in CORSIKA.
This atmospheric model corresponds to the average April South Pole atmosphere at IceTop.
The chosen observation level is the approximate height above sea level of IceTop, corresponding to an atmospheric depth of about 690\,$\text{g} \: \text{cm}^{-2}$.
This atmospheric depth is reached prior to shower maximum for some simulated showers, preventing these air showers from fully developing within the simulations.
To ensure reasonable \xmax estimation for such clipped showers, simulations of air showers with \xmax near (within 30\,$\text{g} \: \text{cm}^{-2}$) or below ground were rerun with the CONEX~\cite{Pierog:2004re, Bergmann:2006yz} option and the observation level set to sea level.
This option allows the development of the air shower to be modeled by solving cascade equations for the separate particle types instead of using the typical CORSIKA Monte Carlo methods; however, from the rerun simulations, only information of the air-shower longitudinal profile is used in this analysis, and the other observables are taken at the observer level from the original simulation.
All IceCube air-shower simulations have high- and low-energy hadronic interactions modeled with Sibyll 2.3d~\cite{Riehn:2019jet} and FLUKA~\cite{Ferrari:2005zk, BOHLEN2014211} respectively, whereas electromagnetic interactions were modeled with the EGS4 code~\cite{Nelson:1985ec}.
In addition, all simulations used the CORSIKA thinning algorithm, set to $10^{-6}$, for a significant reduction in the necessary computation time of the simulations.

Simulation libraries are available for proton (p), helium (He), oxygen (O), and iron (Fe) primary cosmic rays with energies ranging from $\log(E / \text{eV})$ = 16.0$-$18.5 and zenith angles ranging from $\sin^{2}(\theta_{\text{zen}})$ = 0.0$-$0.9 that were produced by us for the main purpose of simulations studies for the IceCube-Gen2 surface array~\cite{IceCube-Gen2:2021jce}.
Energies are binned in 25 evenly spaced bins of width $\log(E / \text{eV})$ = 0.1 while zenith angles are binned in nine evenly spaced bins of width $\sin^{2}(\theta_{\text{zen}})$ = 0.1, with each energy and zenith bin combination containing 200 simulated air showers.
For each air shower, the primary energy and zenith angle are chosen at random within the bin limits.
Azimuth angles are chosen randomly.

CORSIKA v7.6400 was used to simulate cosmic-ray air showers at the site of the Pierre Auger Observatory, with the thinning algorithm set at the same energy threshold as for the IceCube simulated showers.
The observation level was set to 1452\,m above sea level, which is the approximate height of the surface detectors at Auger, and the magnetic field strength was set to 19.4\,\textmu T in the geomagnetic North direction and $-$14.1\,\textmu T downwards.
The electromagnetic and low-energy hadronic interactions were modeled using EGS4 and FLUKA respectively; however, the high-energy hadronic interactions were modeled using Sibyll 2.3c~\cite{Riehn:2017mfm}.
The differences for mass sensitive observables between Sibyll 2.3c and Sibyll 2.3d are described in Ref.~\cite{Riehn:2019jet}, although, as no quantitative comparisons between the observatory sites are performed, these differences will not impact our results.
Furthermore, the differences between CORSIKA versions v7.6400 and v7.7401 are minuscule and non-important for this analysis.

For the Pierre Auger Observatory, general purpose simulation libraries exist for proton, helium, oxygen, and iron cosmic rays~\cite{PierreAuger:2021jov}.
Energies range from $\log(E / \text{eV})$ = 16.0$-$18.5 separated into five equally spaced bins of width $\log(E / \text{eV})$ = 0.5, distributed with an energy spectrum of $dN / dE \propto E^{-1}$.
Each energy bin contains simulated air showers for atmospheric models 18, 20, 25, and 26 within CORSIKA.
These models represent four different months (Jan., Mar., Aug., and Sep. respectively) at Malarg\"ue, Argentina, the site of the observatory.
Each month and energy bin combination contains 1250 air-shower simulations, with zenith angles chosen between limits of 0$^{\circ}$ to 65$^{\circ}$ distributed with $dN / d\theta \propto \sin\theta \cos\theta$ and azimuth angles chosen at random.
Additional air-shower simulations at the site of Auger exist for different energies, zenith angles, and high-energy hadronic interaction models and are described briefly in~\cite{PierreAuger:2021jov}.
These additional simulations were not used in this study to remain consistent with the IceCube simulation library (see Appendix~\ref{appen:additional_auger} for the results of these additional simulations in the context of this analysis).

Table~\ref{tab:simulation_data_table} lists the total number of simulated air showers for different observatories and primary particles.

\begin{table}
    \caption{Air-shower simulation libraries used in this analysis, all within an energy range of $\log(E / \text{eV})$ = 16.0$-$18.5. The number of simulated showers is approximate as some simulated events contain broken simulation files.} \label{tab:simulation_data_table}
    \vspace*{.5\baselineskip}
    \centering
    \begin{ruledtabular}
    \begin{tabular}{ c c c c }
        Location & Primary & Zenith & Number of Showers \\
        \colrule
         & & & \\[-1.0em]
        IceCube & Proton & 0$^{\circ}-$71.6$^{\circ}$ & 45000 \\
        IceCube & Helium & 0$^{\circ}-$71.6$^{\circ}$ & 45000 \\
        IceCube & Oxygen & 0$^{\circ}-$71.6$^{\circ}$ & 45000 \\
        IceCube & Iron & 0$^{\circ}-$71.6$^{\circ}$ & 45000 \\
        Auger & Proton & 0$^{\circ}-$65$^{\circ}$ & 25000 \\
        Auger & Helium & 0$^{\circ}-$65$^{\circ}$ & 25000 \\
        Auger & Oxygen & 0$^{\circ}-$65$^{\circ}$ & 25000 \\
        Auger & Iron & 0$^{\circ}-$65$^{\circ}$ & 25000 \\
    \end{tabular}
    \end{ruledtabular}
\end{table}

From CORSIKA simulations, knowledge of the shape of the electromagnetic shower profile (\xmax, $L$, $R$), the total muon number at ground above an energy threshold (referred to as $N_{\mu}$), the number of electromagnetic particles at \xmax ($N_{\text{e, max}}$), and the number of electromagnetic particles at ground ($N_{\text{e}}$) are obtained.
The photon longitudinal profile of an air shower follows the same form as electron and positron longitudinal profiles, therefore photons are not included in this study, and the electromagnetic particle numbers are determined as the sum of electrons and positrons.
The number of electromagnetic particles at \xmax, $N_{\text{e, max}}$, can be accessed in experiments, e.g., by fluorescence or radio detectors, and serves as an energy proxy to scale air-shower observables.
The number of electromagnetic particles at ground, $N_{\text{e}}$, can be accessed in experiments, e.g., by particle detector arrays, and is used to calculate the electron-muon ratio at ground ($R_{\text{e}/\mu}$). 

\xmax is determined from the CORSIKA simulations by fitting the air-shower electromagnetic longitudinal profile to a Gaisser-Hillas function~\cite{1977ICRCGaisser}, parameterized in terms of $X^{\prime} = X - $\xmax, $L = \sqrt{|X_{0}^{\prime} \lambda|}$, and $R = \sqrt{\lambda / |X_{0}^{\prime}|}$, as

\vspace{-10pt}

\begin{equation}
    N^{\prime} = \left(1 + \frac{R X^{\prime}}{L} \right)^{R^{-2}} \exp{\left( - \frac{X^{\prime}}{L R} \right)}, \label{eq:long_profile_fit}
\end{equation}

\noindent
where $N^{\prime} = N / N_{\text{max}}$ and $X_{0}^{\prime} = X_{0} - $\xmax~\cite{Andringa:2011zz}.
Using this parametrization, $L$ represents a characteristic width of the shower profile while $R$ is related to the asymmetry of the shower profile.
Previous studies hint at the possible mass sensitivity of $L$~\cite{Buitink:2021pkz} and $R$~\cite{Andringa:2011zz} in the contexts of the ultra-dense SKA-low array of radio antennas or precise air-fluorescence observations of ultra-high-energy air showers~\cite{PierreAuger:2018gfc}. 
Appendix~\ref{appen:fit_compare} explains the choice of the parameterized Gaisser-Hillas fit over a typical Gaisser-Hillas fit.

To ensure precise \xmax values, the normalized particle numbers were weighted with respect to their Poissonian fluctuations, which applied more importance to fitting the profile peaks.
For all analyses presented within this paper, only simulated events with fit uncertainties $\sigma_{\text{\xmax}} <$ 5\,$\text{g} \: \text{cm}^{-2}$, $\sigma_{L} <$ 5\,$\text{g} \: \text{cm}^{-2}$, and $\sigma_{R} <$ 0.05 are included.
This removes from the analysis all simulated air showers with anomalous shower profiles, resulting in the removal of 2.2\% (0.8\%) of all proton showers and 0.1\% (0.8\%) of all iron showers at the IceCube (Auger) location.
These are similar to the estimates of anomalous profiles presented in~\cite{Baus:2011kc} and allow for precise knowledge of \xmax, $L$, and $R$ for the simulated events.


\section{Mass Sensitivity of Air-Shower Observables at the IceCube Site} \label{sec:icecube}

The mass sensitivity of all observables obtained from CORSIKA, as described in Section~\ref{sec:simulations}, is studied for air showers simulated at the location of the IceCube Neutrino Observatory.
IceCube has a deep in-ice array of optical detectors in addition to the IceTop surface array of ice-Cherenkov detectors.
The in-ice array is sensitive to muons from air showers with energies above $\mathcal{O}(300$\,GeV) (see Ref.~\cite{IceCube:2013dkx} for a description of the in-ice energy reconstruction at IceCube), motivating the study of a high-energy muon observable specific to the geometry and detector constraints of the in-ice detectors, along with a low-energy muon observable reconstructable by the surface detectors.
Vertical muons from air showers need a minimum energy of 273\,GeV to reach the top of the in-ice detector array, where this energy threshold increases with zenith angle~\cite{IceCube:2015wro}.
For this analysis, the high-energy muon observable refers to $N_{\mu}$ with a fixed energy cut of 500\,GeV as this is above the energy threshold of detectable muons by the deep in-ice array of IceCube detectors, except for highly inclined showers.

The detector signal of surface-particle detector arrays is dominated by electromagnetic particles, complicating measurements of the muon signal for individual events unless detectors have different responses for electromagnetic and muon particles.
With the current IceTop instrumentation, the mean muon density at a given distance to the shower axis was only determined as a statistical average over many air showers~\cite{IceCubeCollaboration:2022tla}.
However, the planned addition of scintillation particle detectors for IceCube-Gen2, in tandem with the current IceTop tanks, may allow for measurements of the low-energy muonic component of individual air showers.
To represent this low-energy surface muon observable, the total muon number within a ground-based annulus is investigated for annuli at equally spaced distances from the air-shower axis.
Annuli at greater distances from the shower axis will hold more importance, based on the geometry and detector constraints of typical surface-particle detector arrays.

\begin{figure}[!b]
\centering
\includegraphics[trim={0in 0in 0in 0in}, clip, width=0.48\textwidth]{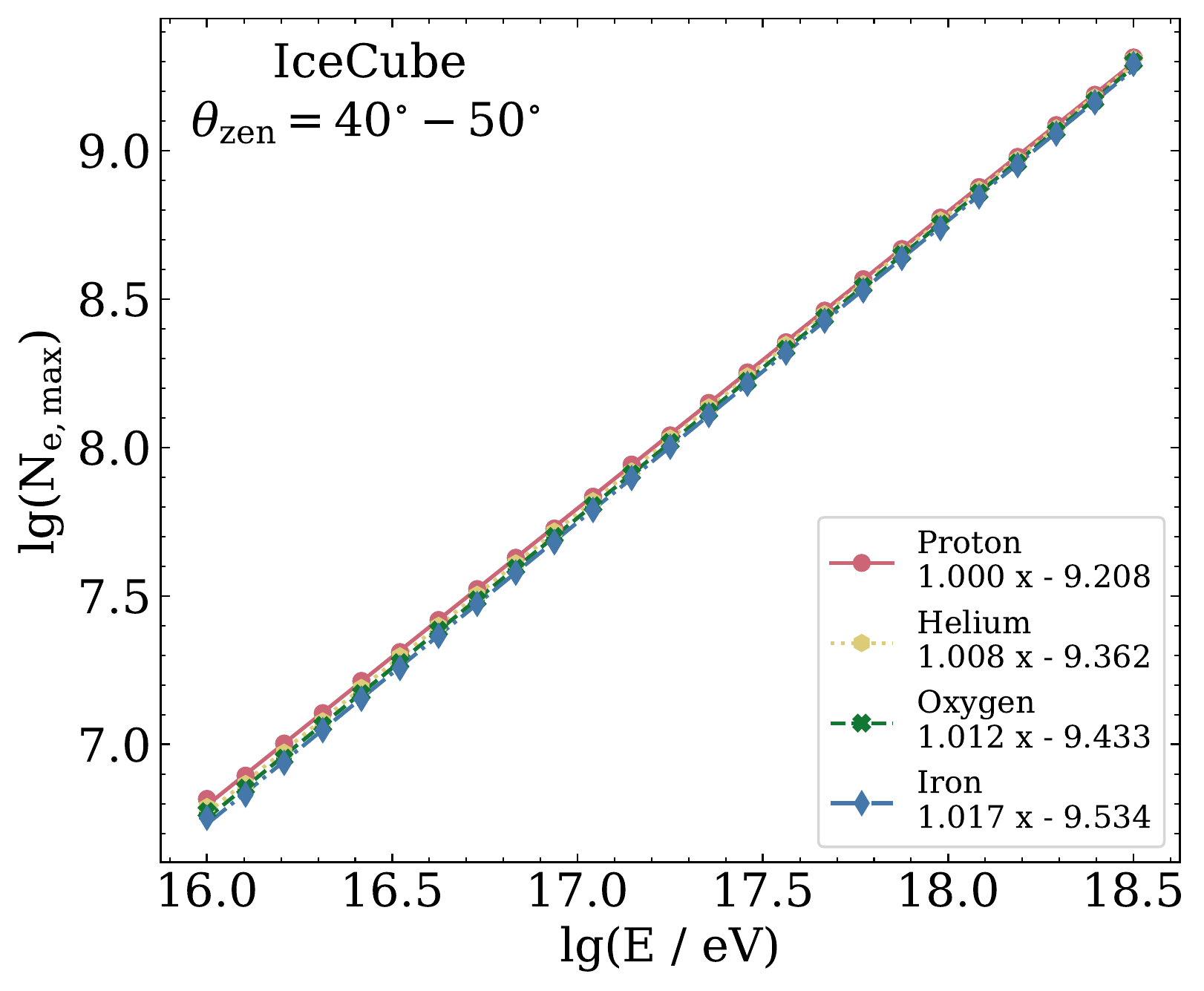}
\caption{Scaling in log-log space of the energy reference observable, $N_{\text{e, max}}$, with respect to simulated air-shower energies at the IceCube Neutrino Observatory.}
\label{fig:nEM_vs_energy}
\end{figure}

\begin{figure}[!t]
\centering
\includegraphics[trim={0in 0in 0in 0in}, clip, width=0.45\textwidth]{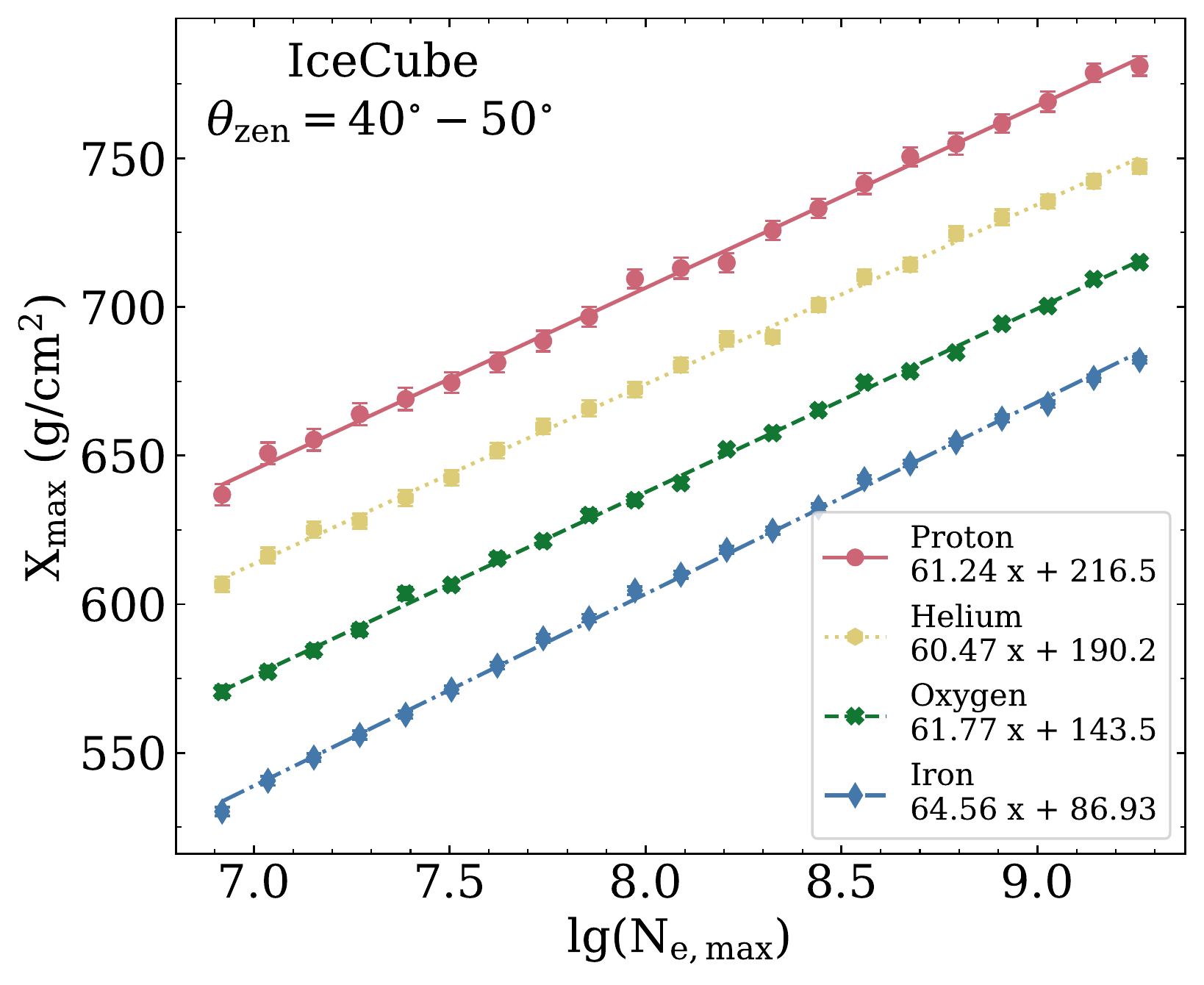}
\includegraphics[trim={0in 0in 0in 0in}, clip, width=0.45\textwidth]{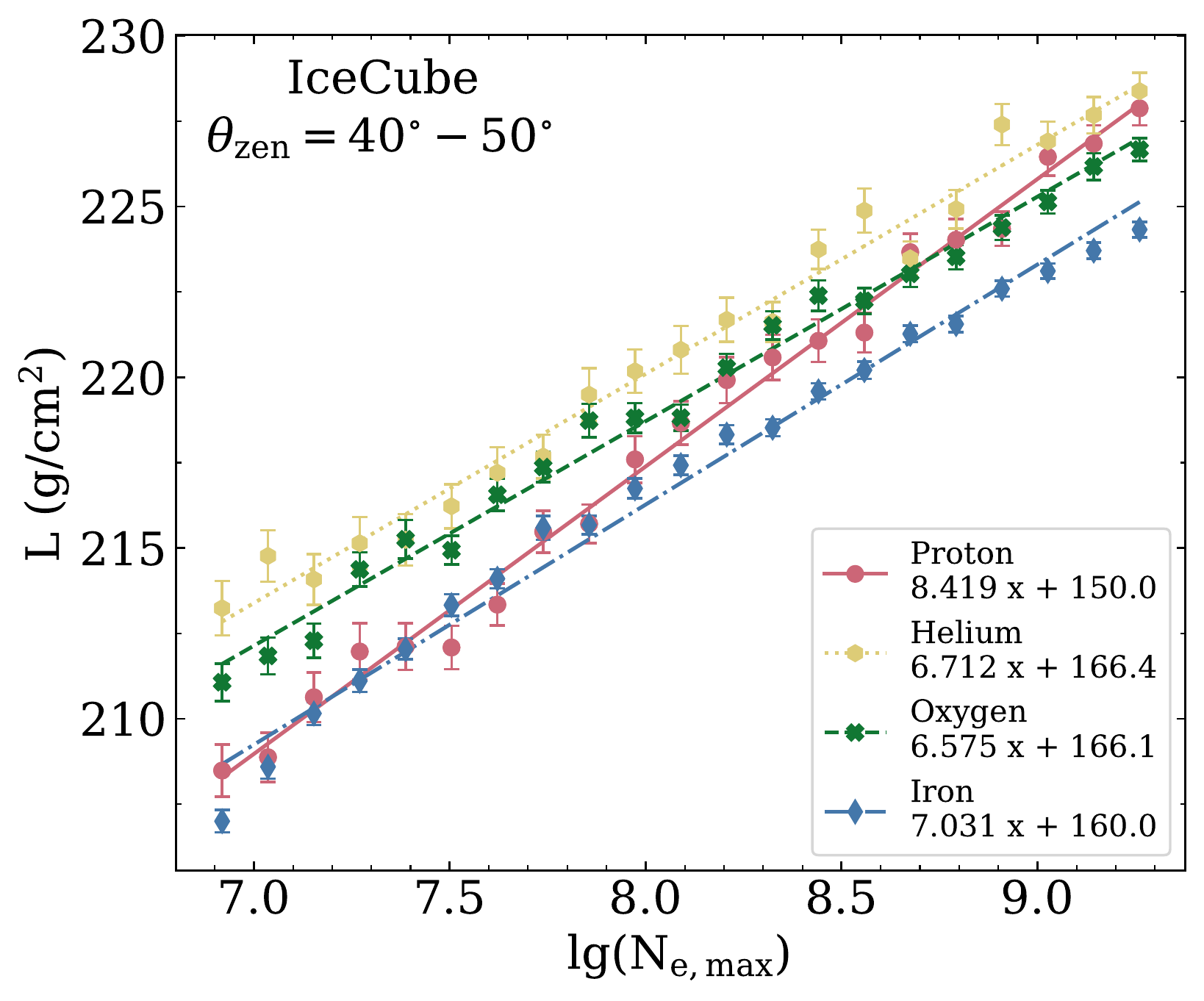}
\includegraphics[trim={0in 0in 0in 0in}, clip, width=0.45\textwidth]{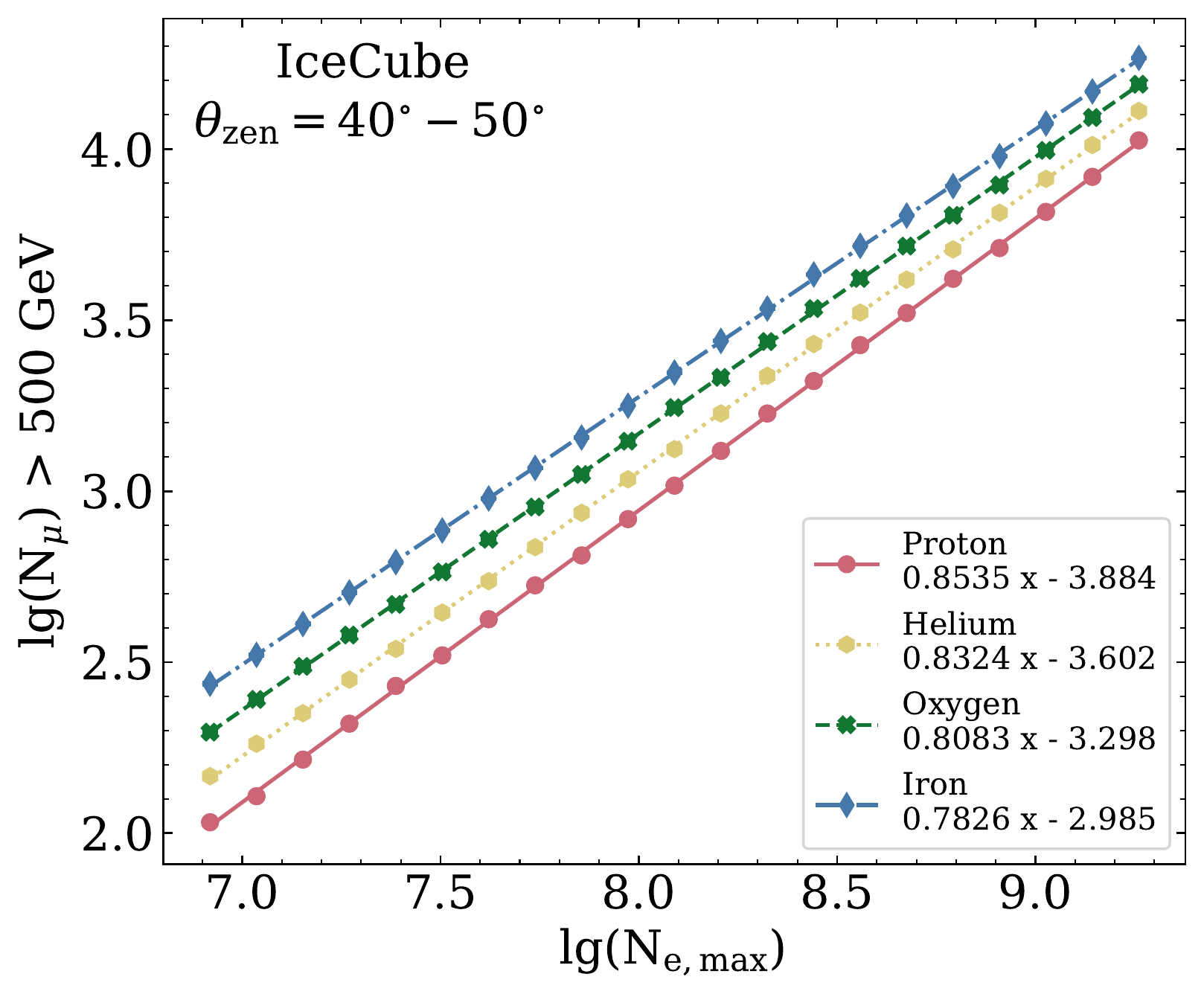}
\caption{Scaling of the observables, \xmax (top), $L$ (middle), and high-energy $\log(N_{\mu})$ (bottom), with respect to the energy reference, $\log(N_{\text{e, max}})$. All scalings are for air showers simulated at the site of the IceCube Neutrino Observatory for a zenith angle range of $\theta_{\text{zen}}$ = 40$^{\circ}-$50$^{\circ}$.}
\label{fig:IC_observable_scaling}
\end{figure}

\subsection{Scaling Corrections for Air-Shower Observables} \label{sec:scaling}

Air-shower observables studied in this analysis must be corrected for their energy dependence, as observable distributions will be smeared from binning simulated air showers with respect to their energy.
The energy of an air shower is not directly observable by current detectors, although the size of the electromagnetic component ($N_{\text{e, max}}$) is used as an energy reference because it is directly proportional to the air-shower energy and can be reconstructed from air-shower observations by radio antennas~\cite{Schroder:2016hrv, Huege:2016veh} or fluorescence detectors.
The dependence of $N_{\text{e, max}}$ on primary energy is depicted in Fig.~\ref{fig:nEM_vs_energy} while the scaling of the observables, \xmax, $L$, and high-energy $N_{\mu}$, with respect to the energy reference are depicted in Fig.~\ref{fig:IC_observable_scaling}.
All air-shower observables were scaled with respect to the energy reference observable, although Fig.~\ref{fig:IC_observable_scaling} only shows the scaling of \xmax, $L$, and high-energy $N_{\mu}$ as examples.
For both figures, the uncertainty in the average observable value for each bin is included, yet in most cases is too small to be discernible.

The zenith range of $\theta_{\text{zen}}$ = 40$^{\circ}-$50$^{\circ}$ was chosen as an example due to the efficiency of the radio antennas at the South Pole in this range~\cite{IceCube:2021qnf} and the importance of this range for the IceCube-Gen2 surface array.
In addition, Fig.~\ref{fig:IC_observable_scaling} has an event number cut of 300 applied to the $N_{\text{e, max}}$ bins to ensure good statistics for all four primaries and full coverage of the electron number at \xmax phase space, the spread in which is caused by shower-to-shower fluctuations.
This event number cut removes bins to the left and right of the energy reference range shown in Fig.~\ref{fig:IC_observable_scaling}, as they do not follow the typical trendlines shown due to their limited phase space coverage.
Removing the bins with limited statistics allows the trendlines to accurately represent the observable scalings, within the energy range used in this analysis.

The average slope of the linear fits to all four primaries was calculated for all observables.
Distinct from all other observables is the scaling of $L$, with the helium trendline raised above that of all other primaries and the proton trendline exhibiting a much steeper slope compared to the other primaries.
However, because the scaling correction of $L$ has no significant effect on the mass separation results shown in this paper, we use the average slope of all four primaries also for $L$.
As in real experiments the true energy is unknown, therefore the observables were corrected for $N_{\text{e, max}}$ instead: $0.01$ was then added to this average slope to correct for the weak deviation of $N_{\text{e, max}}$ from an exact linear energy scaling, as 1.01 is the average of the linear fit slopes shown in Fig.~\ref{fig:nEM_vs_energy}.
Consequently, the corrections to \xmax, $L$, $R$, and $N_{\text{e}}$ are

\vspace{-10pt}

\begin{equation}
    X_{\text{max}} = X_{\text{max, true}} - 62.0 \times \log \left( \frac{N_{\text{e, max}}}{N_{\text{EeV}}} \right), \label{eq:xmax_correction}
\end{equation}

\vspace{-10pt}

\begin{equation}
    L = L_{\text{true}} - 7.19 \times \log \left( \frac{N_{\text{e, max}}}{N_{\text{EeV}}} \right), \label{eq:L_correction}
\end{equation}

\vspace{-10pt}

\begin{equation}
    R = R_{\text{true}} + 0.02 \times \log \left( \frac{N_{\text{e, max}}}{N_{\text{EeV}}} \right), \label{eq:R_correction}
\end{equation}

\vspace{-10pt}

\begin{equation}
    N_{\text{e}} = \frac{N_{\text{e, true}}}{\left( \frac{N_{\text{e, max}}}{N_{\text{EeV}}} \right)^{1.14}} \label{eq:em_num_correction}
\end{equation}

\noindent
where $N_{\text{EeV}}$ is the average number of electrons at \xmax for simulated air showers with energies of 1\,EeV.
Corrections to the observables $N_{\mu}$ and $R_{\text{e}/\mu}$ = $\log(N_{\text{e}} / N_{\mu})$ for high-energy muons are, 

\vspace{-10pt}

\begin{equation}
    N_{\mu} = \frac{N_{\mu, \text{true}}}{\left( \frac{N_{\text{e, max}}}{N_{\text{EeV}}} \right)^{0.83}}, \label{eq:muon_num_correction}
\end{equation}

\vspace{-10pt}

\begin{equation}
    R_{\text{e}/\mu} = R_{\text{e}/\mu, \text{true}} - 0.32 \times \log \left( \frac{N_{\text{e, max}}}{N_{\text{EeV}}} \right). \label{eq:ratio_correction}
\end{equation}

For the low-energy surface muon observables, annuli of 50\,m width were investigated with the first and last annuli respectively ranging from 0$-$50\,m and 950$-$1000\,m from the air-shower axis.
Within each annulus, the $N_{\mu}$ and $N_{\text{e}}$ observables were used to calculate the electron-muon ratio of the annulus.
The scalings of each muon annulus follow closely the scaling of the total muon number at ground, whereas the scaling of the electron-muon ratios varies depending on the distance to the shower axis.
Hence, the muon annulus scaling factor was calculated from the scaling of the total muon number at ground, while the annulus electron-muon ratio scaling was handled on a case-by-case basis.
The average slope from the linear fits to all four primaries was calculated and added to the energy dependence of the $N_{\text{e, max}}$ observable.
Corrections to the low-energy annulus $N_{\mu}$ and $R_{\text{e}/\mu}$ observables are

\vspace{-10pt}

\begin{equation}
    N_{\mu, \text{ann}} = \frac{N_{\mu, \text{ann-true}}}{\left( \frac{N_{\text{e, max}}}{N_{\text{EeV}}} \right)^{0.94}}, \label{eq:muon_annulus_correction}
\end{equation}

\vspace{-10pt}

\begin{equation}
    R_{\text{e}/\mu, \text{ann}} = R_{\text{e}/\mu, \text{ann-true}} - (m + 0.01) \times \log \left( \frac{N_{\text{e, max}}}{N_{\text{EeV}}} \right) \label{eq:ratio_annulus_correction}
\end{equation}

\noindent
where $m$ is the slope of the linear fit, which varies between 0.09 and 0.20 depending on the annulus distance to the shower axis.
Based on IceTop observations, the muon densities at both 600\,m and 800\,m from the shower axis can be reconstructed at the South Pole~\cite{IceCubeCollaboration:2022tla}.
Therefore, the 800$-$850\,m annulus was chosen for use in the mass sensitivity analysis.
This chosen annulus requires a value of $m$ = 0.09 in the scaling of the annulus electron-muon ratio, given by Eq.~\ref{eq:ratio_annulus_correction}.
Further explanation for the choice of annulus is provided in Appendix~\ref{appen:annulus_FOM}.

\begin{figure}[!t]
\centering
\includegraphics[trim={0in 0in 0in 0in}, clip, width=0.47\textwidth]{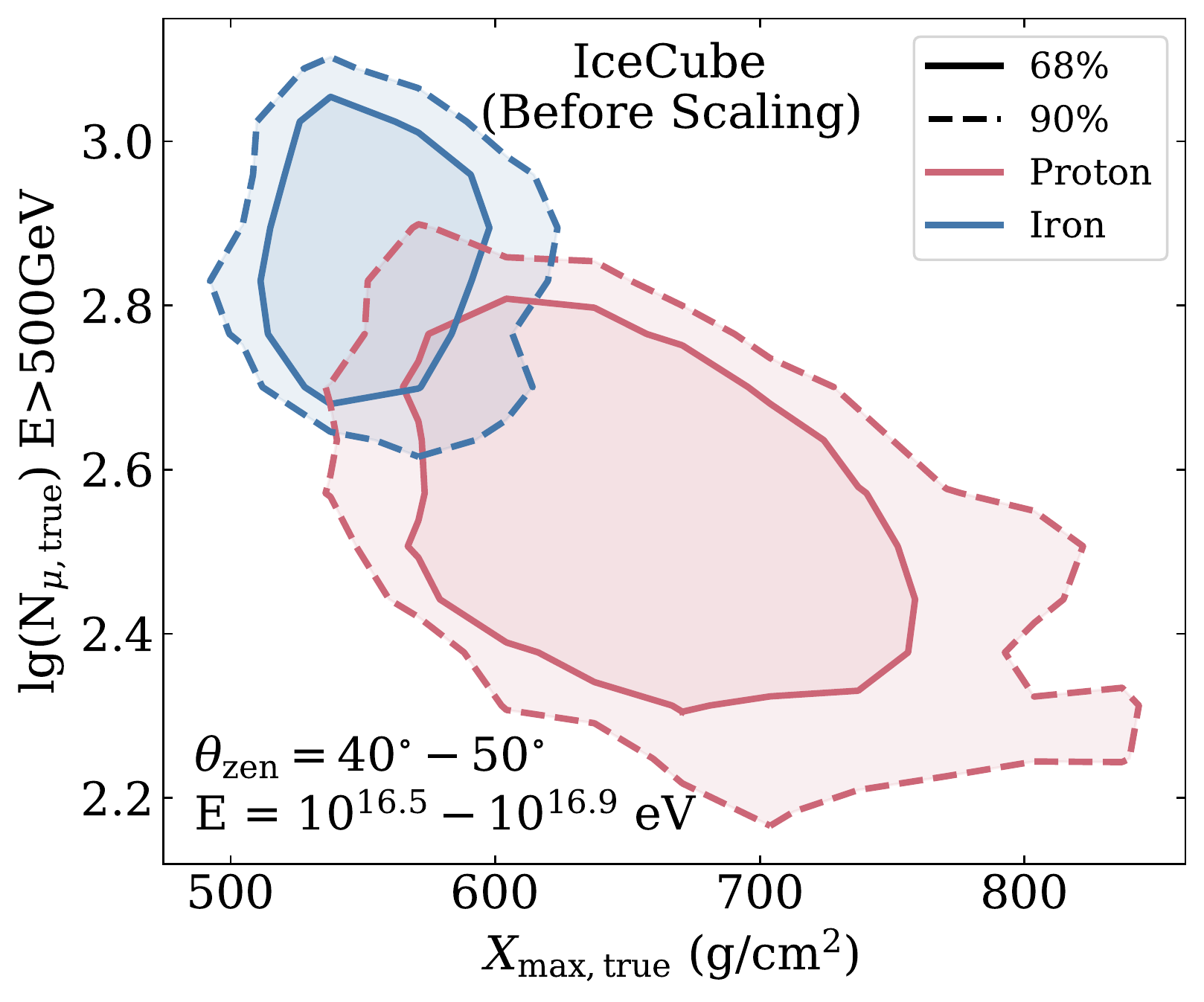}
\includegraphics[trim={0in 0in 0in 0in}, clip, width=0.47\textwidth]{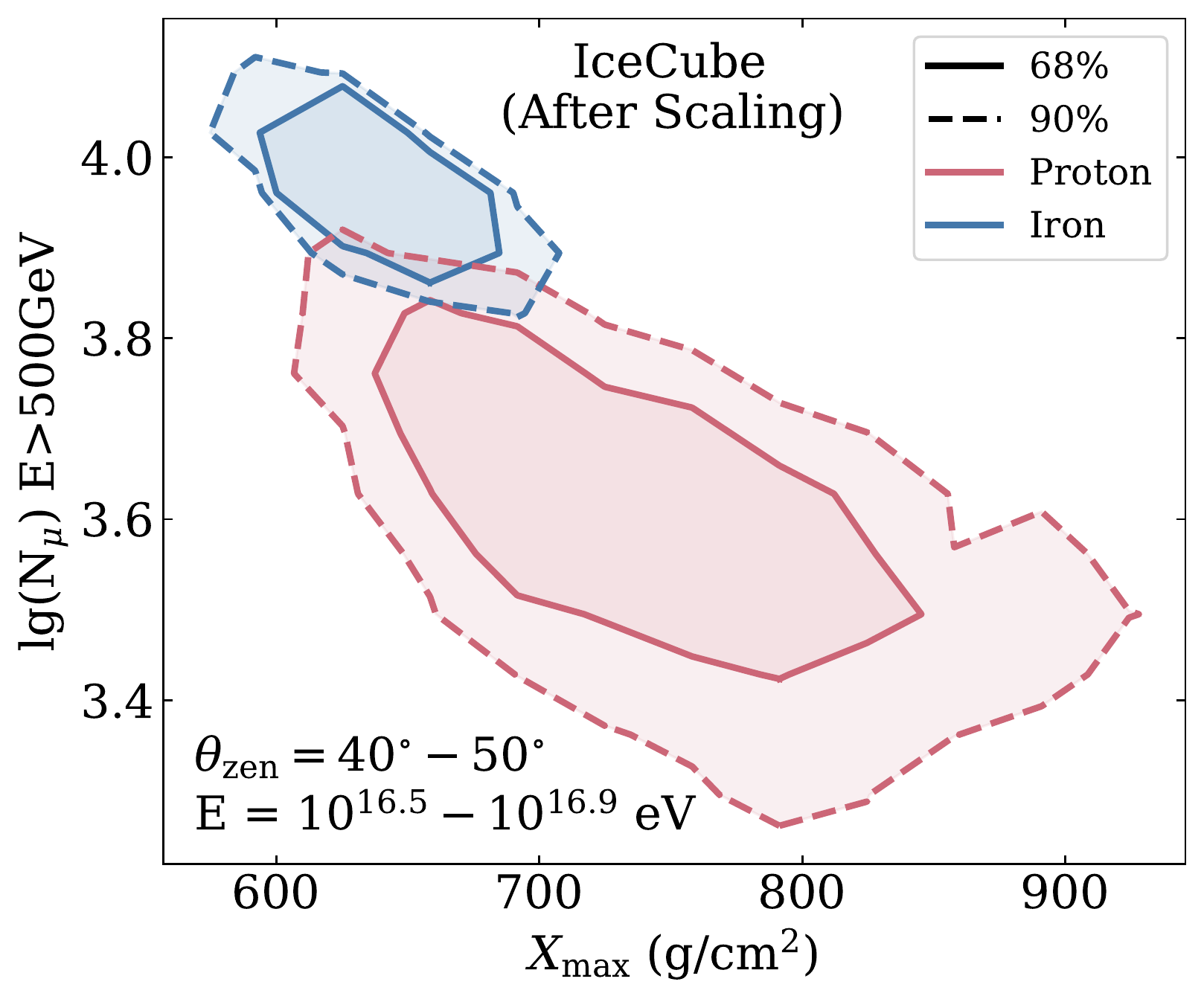}
\caption{Two-dimensional contour plots of the proton-iron distributions for the high-energy muon and \xmax observables before (top) and after (bottom) applying the energy-scaling correction described in the text.}
\label{fig:IC_contours_scaling_example}
\end{figure}

Scaling the air-shower observables removes the energy dependence of these observables as seen in the simulations, allowing for the use of wide energy bins in our analysis without risk of smearing the observable distributions.
As a consequence, this scaling also increases the mass separation of the individual observables, as illustrated by the proton-iron contour plots in Fig.~\ref{fig:IC_contours_scaling_example}.
Only proton and iron distributions are shown in this figure for clarity, highlighting the effect of scaling on the high-energy $N_{\mu}$ and \xmax observables.

\subsection{Mass Sensitivity of Observables Accounting for Shower-to-Shower Fluctuations} \label{sec:sensitivity_w_fluctuations}

\begin{figure}[!t]
\centering
\includegraphics[trim={0in 0in 0in 0in}, clip, width=0.47\textwidth]{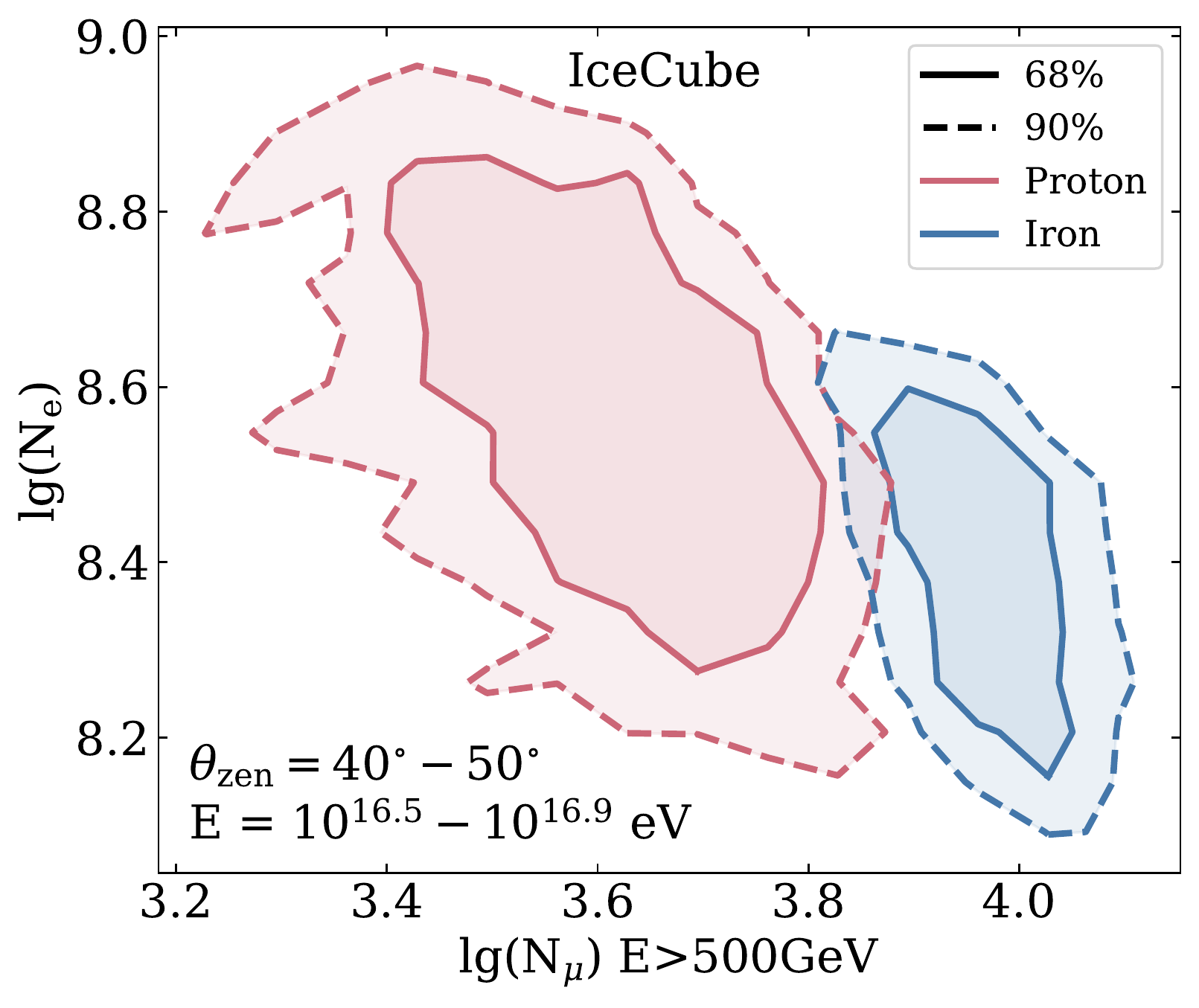}
\includegraphics[trim={0in 0in 0in 0in}, clip, width=0.47\textwidth]{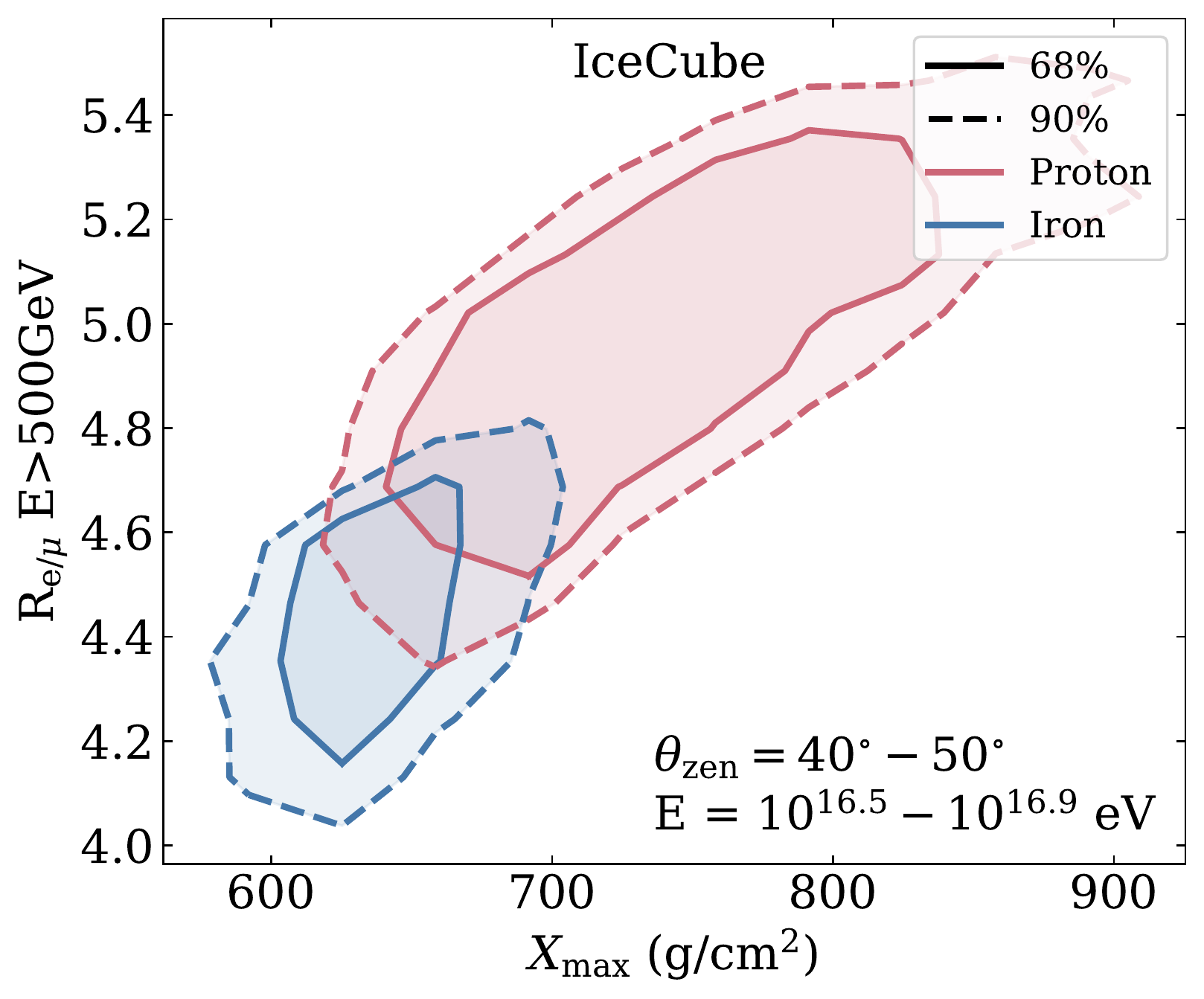}
\caption{Two-dimensional contour plots of the proton-iron distributions for the $N_{\text{e}}$ and high-energy muon observables at ground level (top); and their ratio $R_{\text{e}/\mu}$ at ground level and \xmax (bottom). All observables are scaled with respect to the energy reference observable.}
\label{fig:IC_contours}
\end{figure}

The mean mass separation of individual observables is visible in Fig.~\ref{fig:IC_observable_scaling}, but to study the event-by-event mass separation also the statistical shower-to-shower fluctuations of the observables are important in addition to the separation of the mean values.
Therefore, exact knowledge of mass sensitive observables is taken from air-shower simulations on a per-event basis and used to construct the two-dimensional distributions depicted as contour plots in Fig.~\ref{fig:IC_contours}, where a primary energy range of 10$^{16.5}-$10$^{16.9}$\,eV and zenith range of 40$^{\circ}-$50$^{\circ}$ are exemplary shown.
The width of the energy and zenith bins has been chosen narrow enough to see trends over energy and zenith angles, on the one hand, and to provide sufficient statistics in each bin, on the other hand.

The combination of high-energy muon number and \xmax for the same energy and zenith ranges was already presented in Fig.~\ref{fig:IC_contours_scaling_example}.
The minimal overlap of the proton and iron contours clearly illustrates that the $N_{\text{e}}$ and high-energy $N_{\mu}$ combination provides a good mass separation, mostly due to the separation power of the high-energy muons.
Helium and oxygen distributions are excluded here for clarity, as in Fig.~\ref{fig:IC_contours_scaling_example}.

A Fisher linear discriminant analysis was performed to find the line of best separation, known as the Fisher axis, between the observable distributions of the proton and iron cosmic rays.
Essentially, this linear discriminant analysis reduces the dimensionality of the multivariate analysis to a single dimension without the loss of separation information between variables.
From the Fisher analysis, a figure of merit (FOM) value, defined as

\begin{figure}[!t]
\centering
\includegraphics[trim={0in 0in 0in 0in}, clip, width=0.47\textwidth]{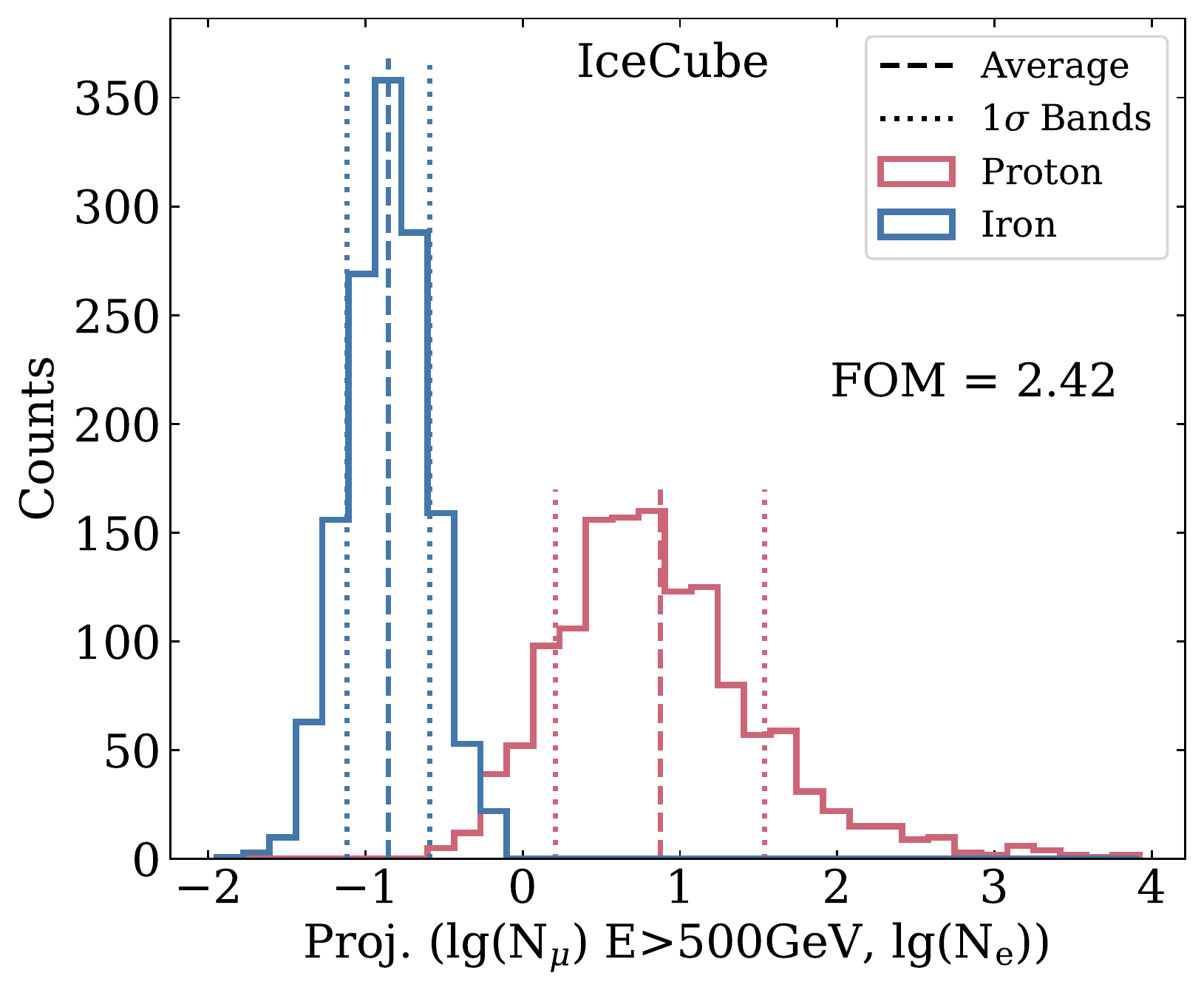}
\includegraphics[trim={0in 0in 0in 0in}, clip, width=0.47\textwidth]{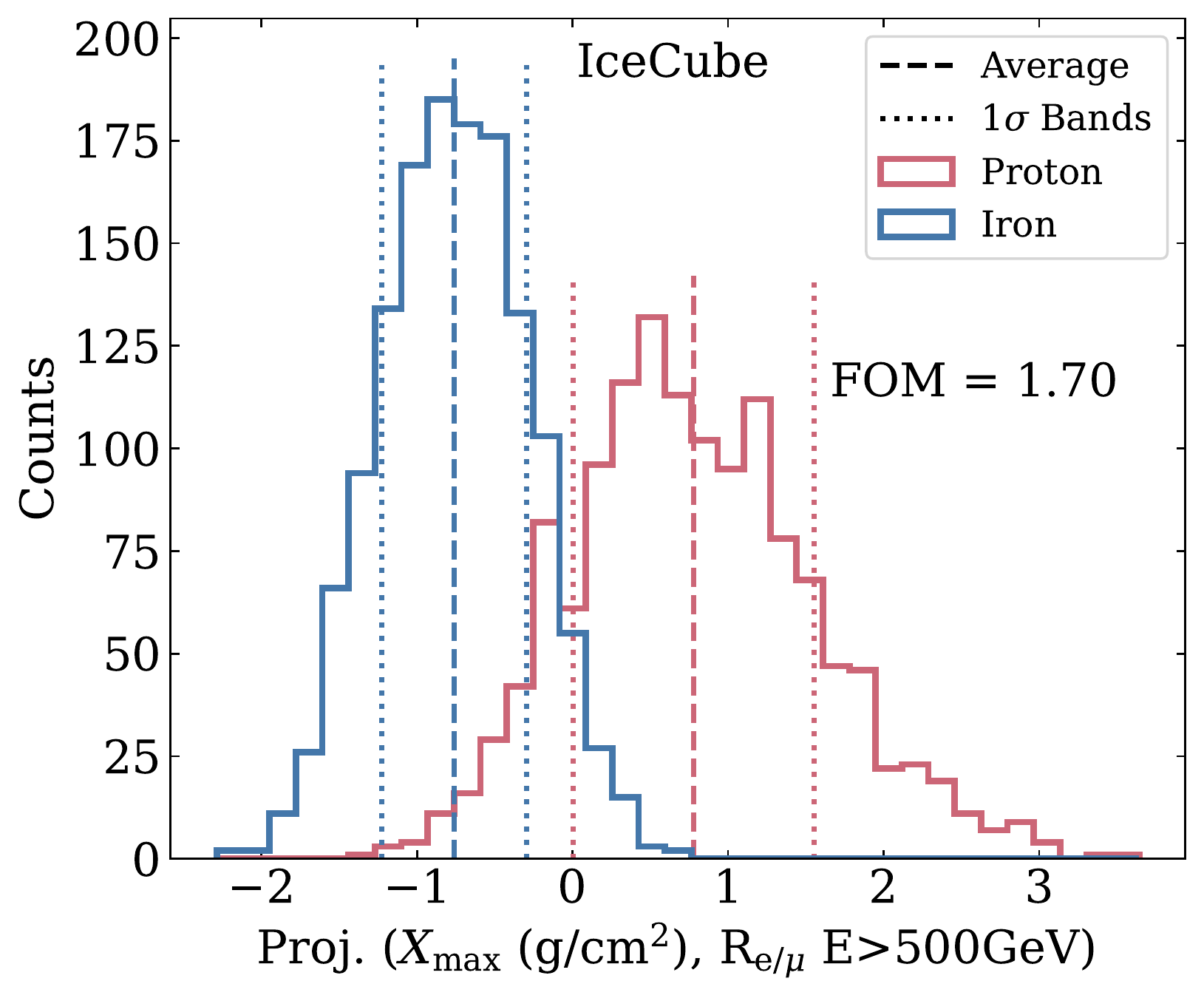}
\caption{Projections onto the one-dimensional Fisher axis of the observable combinations from Fig.~\ref{fig:IC_contours}. The FOM value introduced in Eq.~\ref{eq:figure_of_merit} is calculated and overlaid for each plot.}
\label{fig:IC_fisher_projections}
\end{figure}

\vspace{-10pt}

\begin{equation}
    \text{FOM} = \frac{| \mu_{\text{p}} - \mu_{\text{Fe}} |}{\sqrt{\sigma_{\text{p}}^{2} + \sigma_{\text{Fe}}^{2}}}, \label{eq:figure_of_merit}
\end{equation}

\noindent
was calculated.
The $\mu$ and $\sigma$ values respectively represent the mean and standard deviation of the primary particle distributions, after being projected onto the one-dimensional Fisher axis.
Hence, the FOM is calculated statistically from knowledge of the observable distributions, but serves as a measure of the mass-separation power on an event-by-event basis.
Separation power between the distributions increases with increasing FOM values, where it is typically held that FOM values larger than 1.5 correspond to distributions which can be reasonably separated.
Projection plots on the one-dimensional Fisher axis, for the same observable combinations from Fig.~\ref{fig:IC_contours}, are presented in Fig.~\ref{fig:IC_fisher_projections}.
FOM values are also calculated for each projection plot to numerically represent the separation power for the observable combinations.
Clearly, the observable combination of $N_{\text{e}}$ and high-energy $N_{\mu}$ yields the better proton-iron separation power, which is in agreement with the smaller overlap in the two-dimensional contour plots shown in Fig.~\ref{fig:IC_contours}.

\begin{figure}[!t]
\centering
\includegraphics[trim={0in 0in 0in 0in}, clip, width=0.47\textwidth]{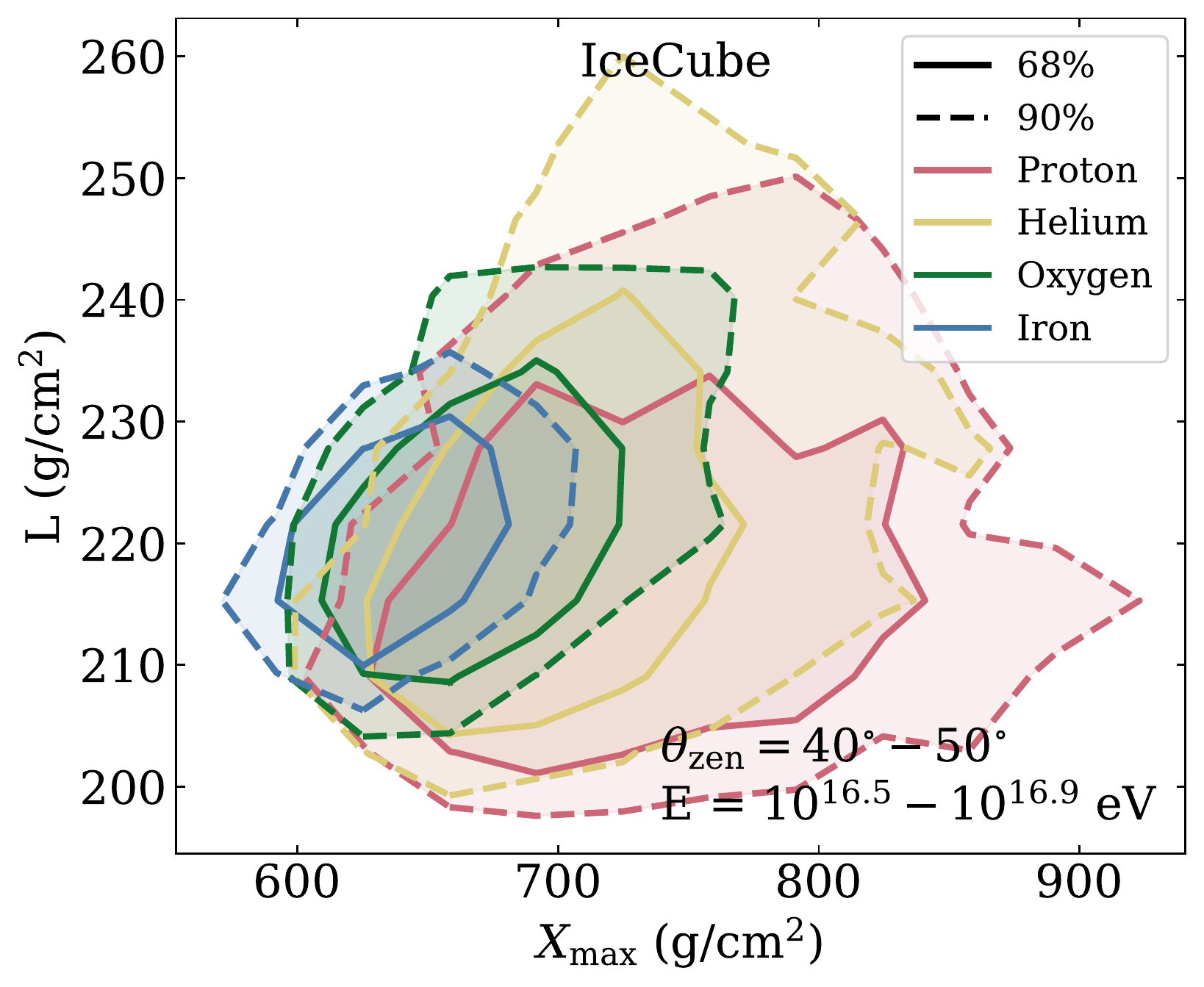}
\caption{Two-dimensional contour plot of the $L$ and \xmax air-shower observables for simulations of all four primaries studied (proton, helium, oxygen, iron). Both the $L$ and \xmax observables are corrected with respect to the energy reference observable from Eqs.~\ref{eq:L_correction} and~\ref{eq:xmax_correction} respectively.}
\label{fig:IC_L_Xmax_contour}
\end{figure}

Motivated by another simulation study which showed that the combination of $L$ and \xmax observables can be used to estimate the helium fraction of primaries if high statistics are available and average observable values are studied~\cite{Buitink:2021pkz}, we had a closer look to the mass separation of these observables (Fig.~\ref{fig:IC_L_Xmax_contour}).
We generally confirm the results of Ref.~\cite{Buitink:2021pkz}, in particular, a visible outcropping of the helium distribution for $L$.
This is caused by helium having the longest distribution tail for the $L$ observable, instead of proton which has the longest tail in the distributions of all other studied shower observables.
Therefore, $L$ might indeed be used to determine the proton-helium ratio in the cosmic-ray flux when measured with sufficient accuracy and high statistics.
Nonetheless, measuring $L$ does not significantly improve the event-by-event mass separation, as made clear by the largely overlapping 68\% contours in Fig.~\ref{fig:IC_L_Xmax_contour}.

\subsection{Event-by-Event Separation of Individual Observables at Different Reconstruction Resolutions} \label{sec:IC_individual_observables}

Important for any air-shower analysis is the resolution at which air-shower observables can be reconstructed.
For a given observable, both the type of ground-based detection technique and detector specific systematics play a role in the reconstruction accuracy and precision.
For IceCube-Gen2, the reconstruction resolution of certain air-shower observables studied in this analysis is not yet fully understood.
However, a range of uncertainties for observable reconstruction can be investigated, where the baseline uncertainties are motivated by the current generation IceCube detectors and reconstruction techniques of detectors similar to the planned upgrades, yet located at other observatories around the world.

\begin{figure*}[!t]
\centering
\includegraphics[trim={0in 0in 0in 0in}, clip, width=0.808\textwidth]{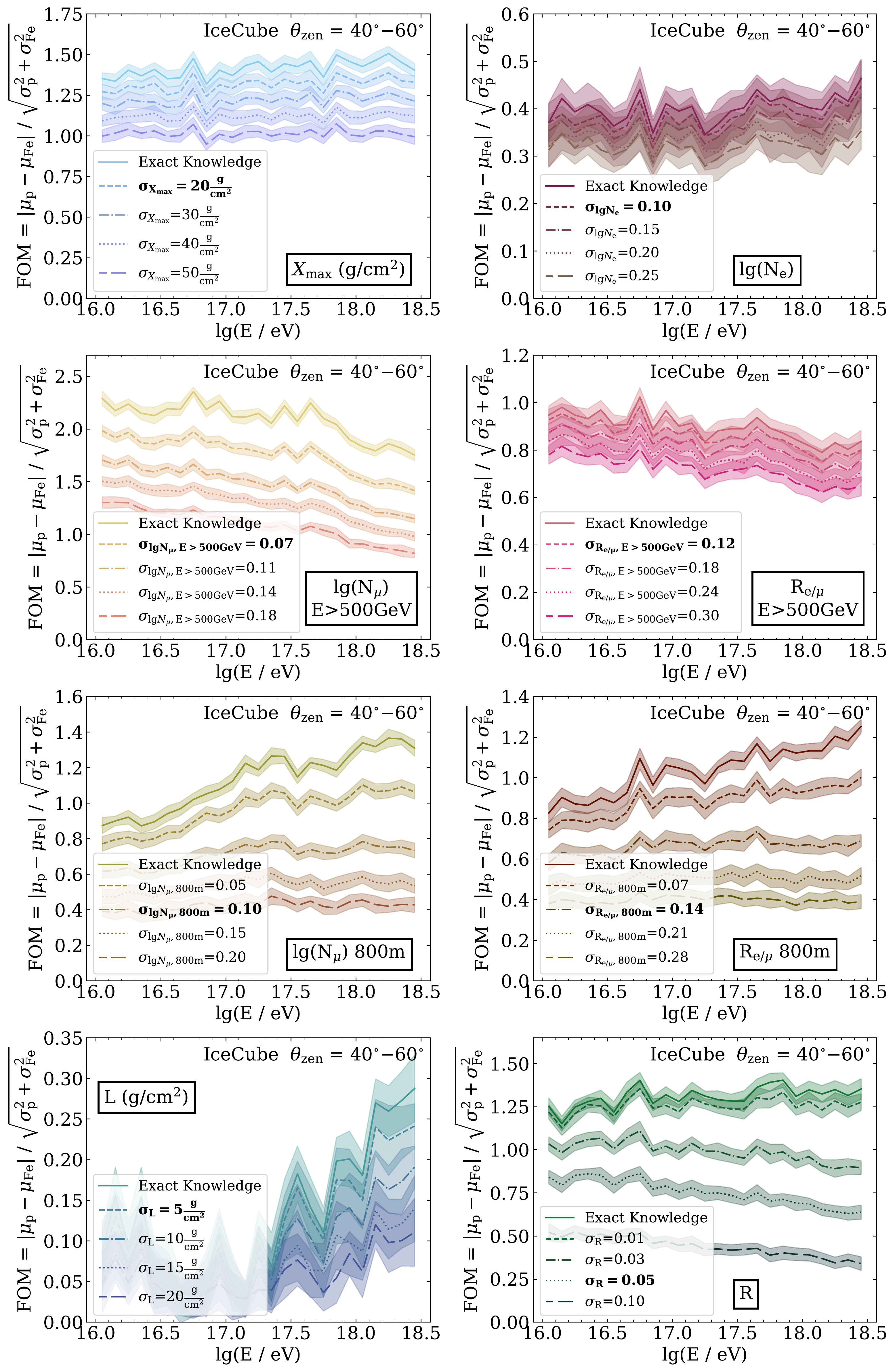}
\caption{Proton-iron separation power plots as a function of air-shower energy for knowledge of a single air-shower observable. A range of reconstruction uncertainties for each observable are assumed and their respective separation power curves are included within the plots, where the assumed baseline uncertainties are in boldface font within the legend. All observables are scaled with respect to the energy reference observable $N_{\text{e, max}}$. Except for the 'exact knowledge' curves, an uncertainty of $\sigma_{N_{\text{e, max}}} = 0.1$ was applied (see text).}
\label{fig:IC_single_observables}
\end{figure*}

For each air-shower observable, proton-iron separation power curves as a function of air-shower energy were calculated using the FOM metric.
To estimate the uncertainty of the FOM, a bootstrapping random sampling with replacement method was performed.
This process was repeated to calculate 30 FOM values, of which the average and standard deviation were taken as the nominal FOM and its 1$\sigma$ uncertainty band.
Each observable had five curves calculated, one with exact knowledge of the scaled observable, one assuming the observable is known within a baseline uncertainty, and three additional curves with reconstruction uncertainties different from the assumed baseline.

The FOM values for all observables were calculated with knowledge of both the observable in question and the energy reference, $N_{\text{e, max}}$.
Except for the FOM values labeled 'exact knowledge', an uncertainty in $N_{\text{e, max}}$ of 10\% was applied, motivated by the order of magnitude of current uncertainties of radio and air-fluorescence energy measurements \cite{PierreAuger:2015hbf,Tunka-Rex:2015zsa,LOPES:2021ipp,PierreAuger:2017ild}.
Separate resolutions for this observable were not investigated.
The baseline \xmax uncertainty is assumed to be 20\,$\text{g} \: \text{cm}^{-2}$, in between the uncertainty from~\cite{Buitink:2014eqa,Bezyazeekov:2018yjw}.
Also, this \xmax uncertainty is near the resolution expected for the full array of elevated radio antennas at IceTop once deployed~\cite{Turcotte:2022}.
The baseline uncertainty for $\log(N_{\text{e}})$ is assumed to be 0.1; as the IceTop signal is dominated by the electromagnetic component and measured close to the shower maximum, the resolution on $N_{\text{e}}$ might be even better~\cite{IceCube:2012nn}, however, the effect of the resolution on the FOM is small, anyway.

The baseline uncertainty of $\log(N_{\mu})$ for a 500\,GeV energy cut is taken to be 0.07 from~\cite{Verpoest:2022ntf} while the uncertainty of $\log(N_{\mu})$ within the 800$-$850\,m annulus is assumed to be 0.1.
Baselines for their respective electron-muon ratios are found by error propagation of the respective individual uncertainties.
Motivated by Ref.~\cite{PierreAuger:2018gfc}, the $R$ baseline uncertainty is assumed to be 0.05.
The $L$ baseline uncertainty is assumed to be 5\,$\text{g} \: \text{cm}^{-2}$ which is lower than the reconstruction uncertainty from~\cite{PierreAuger:2018gfc}; however, the $L$ resolution has little impact on the proton-iron separation power for this observable.
Currently, IceCube does not measure either $R$ or $L$, so the study primarily serves the purpose of whether it is worth adding detection capabilities for these observables if a certain resolution could be obtained.
As shown in Fig.~\ref{fig:IC_single_observables}, the proton-iron separation for certain air-shower observables, such as $\log(N_{\text{e}})$, is not heavily influenced by the observable's reconstruction uncertainty.
Whereas other observables, such as $\log(N_{\mu})$, have a much higher dependence of the proton-iron discrimination power on the assumed observable reconstruction uncertainty.


\section{Mass Sensitivity of Air-Shower Observables at the Auger Site} \label{sec:auger}

To check whether the results obtained for the IceCube site at the South Pole are specific to that particular location or transferable to other sites, we have performed a corresponding study of the event-by-event mass separation also for the site of the Pierre Auger Observatory.
All observables from the previous section were investigated at the Auger location, with the exception of the high-energy ($>$ 500\,GeV) muon observables, as Auger has no detection technique to distinguish these muons from those at lower energies.

The energy reference scaling of these observables for the Auger simulation library was performed separately from the IceCube scaling because environmental factors, such as the altitude of the observation level and the atmospheric model, affect the total muon number within an air shower.
Similar to the energy reference scaling in Section~\ref{sec:scaling}, an event number cut on the $N_{\text{e, max}}$ bins must be applied.
This removes outlying bins at the lowest and highest energies which do not have full phase space coverage, allowing for a continuous energy interval between these outlying bins to be used for the energy reference scaling.
Due to the lower shower statistics, the event number cut for the Auger $N_{\text{e, max}}$ bins is 200, instead of 300 used for the IceCube simulations.

The scaling of the low-energy muon observables, along with the scaling of the energy reference observable with respect to the air-shower energy, follow the same behavior as the IceCube simulated air showers and therefore Eqs.~\ref{eq:muon_annulus_correction} and~\ref{eq:ratio_annulus_correction} will be used to correct the annulus observables for low-energy muons.
However, for corrections to the Auger observables, $m$ varies between 0.10 and 0.23 and the $N_{\text{EeV}}$ normalization factor is 3\% smaller than that for IceCube simulations.
To remain consistent with the analysis of IceCube simulations, the mass sensitivity of muon observables in the 800$-$850\,m annulus was investigated.
This choice of annulus corresponds to a value of $m$ = 0.11 for Eq.~\ref{eq:ratio_annulus_correction}, which is slightly higher than the value used for IceCube simulations.
The mass sensitivity of a low-energy surface muon observable at the site of Auger was previously studied for a fixed air-shower energy~\cite{Holt:2019fnj}, although the low-energy muon observable used in this study is farther from the air-shower axis than in the previous analysis.
In addition, the scaling of the $R$ observable remains consistent between the Auger and IceCube locations, therefore Eq.~\ref{eq:R_correction} was used to correct the Auger $R$ observable.
The Auger \xmax, $L$, and $N_{\text{e}}$ observables have slightly steeper scalings with respect to the energy reference observable than these same scalings at the IceCube site.
Hence, for Auger, the constant terms used to multiply the logarithmic energy reference observable become 62.8 for Eq.~\ref{eq:xmax_correction} and 7.48 for Eq.~\ref{eq:L_correction}, while the exponent used to scale the energy reference observable in Eq.~\ref{eq:em_num_correction} becomes 1.17.

In Section~\ref{sec:sensitivity_w_fluctuations}, the $L$ observable was shown to exhibit a behavior distinct from all other studied observables, as the helium distribution of $L$ had the largest tail of the four primaries.
Qualitatively, the same is observed with $L$ values from air-shower simulations at the Auger site.
A similar outcropping to that seen in Fig.~\ref{fig:IC_L_Xmax_contour} is observed when investigating the mass separation of the \xmax and $L$ observable combination at Auger; however, the helium distinction is not as prominent due to the proton $L$ distribution at Auger having a larger tail than this same distribution at IceCube.

The proton-iron separation power of individual observables, at multiple reconstruction uncertainties, was also studied at the Auger site.
Mass separation for all observables, minus high-energy muons, were studied using the same reconstruction uncertainties from Section~\ref{sec:IC_individual_observables}.
The effect of various reconstruction uncertainties on the observables at Auger is almost identical to the effect on observables at IceCube.
However, particularly for inclined air showers, the low-energy $N_{\mu}$ observable at Auger exhibits increased proton-iron mass sensitivity compared to this observable at IceCube.
Due to the observation level difference between the IceCube and Auger sites, muons at 800\,m from the air-shower axis at IceCube would have propagated to further distances from the shower axis upon reaching the Auger observation level.
Yet, choosing a muon annulus closer to the air-shower axis at IceCube does not alleviate the difference in proton-iron mass separation for this observable between the IceCube and Auger locations, and further studies will be needed about the influence of a particular site and observation altitude on the mass-sensitivity of muon surface detectors.
Apart from the surface muons, the figures showing the results for Auger are similar to those for observables at IceCube and are included in Appendix~\ref{appen:auger_plots}.


\begin{figure*}[!t]
\centering
\includegraphics[trim={0in 0in 0in 0in}, clip, width=0.99\textwidth]{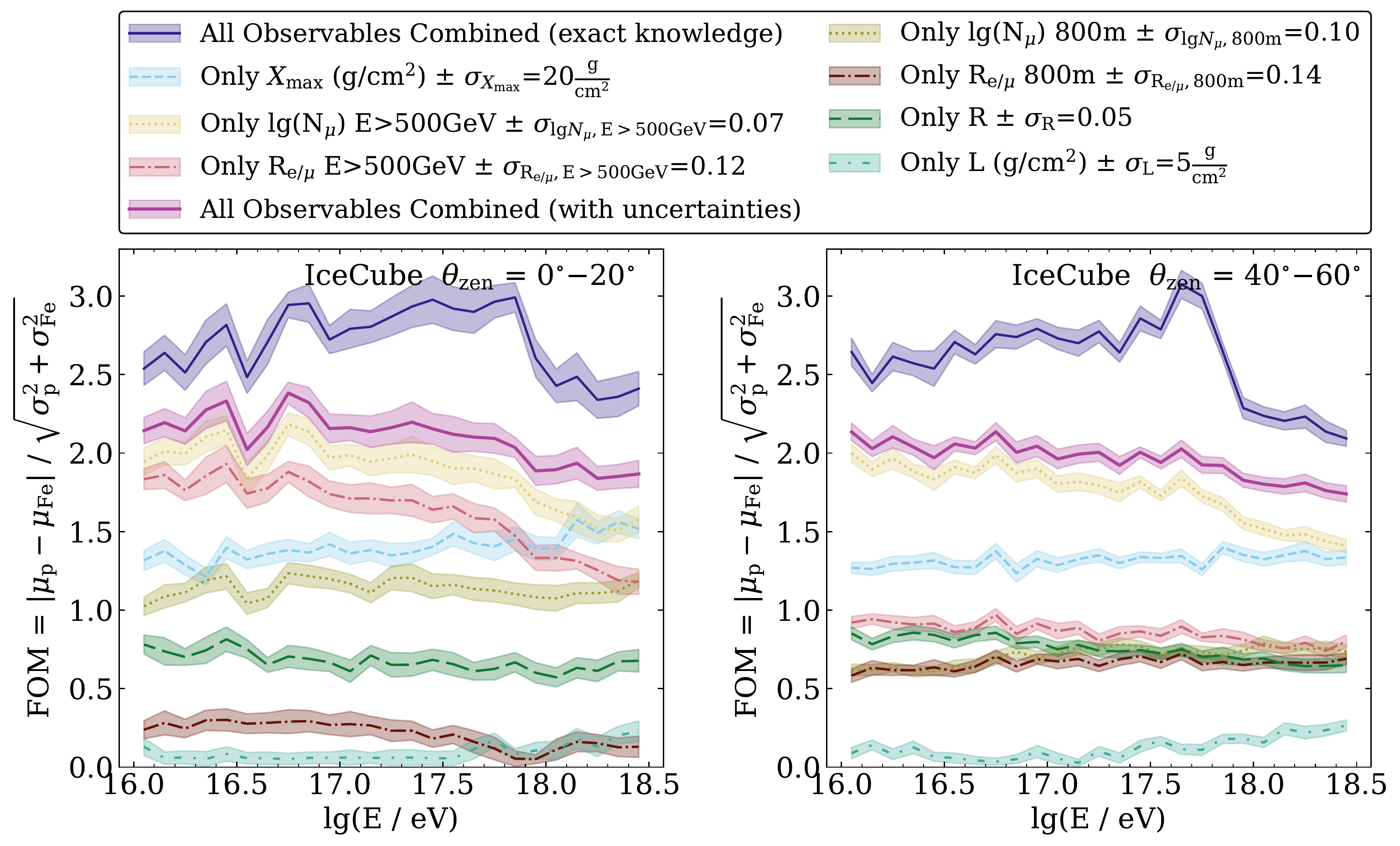}
\vspace{-1.2em}
\includegraphics[trim={0in 0in 0in 0in}, clip, width=0.99\textwidth]{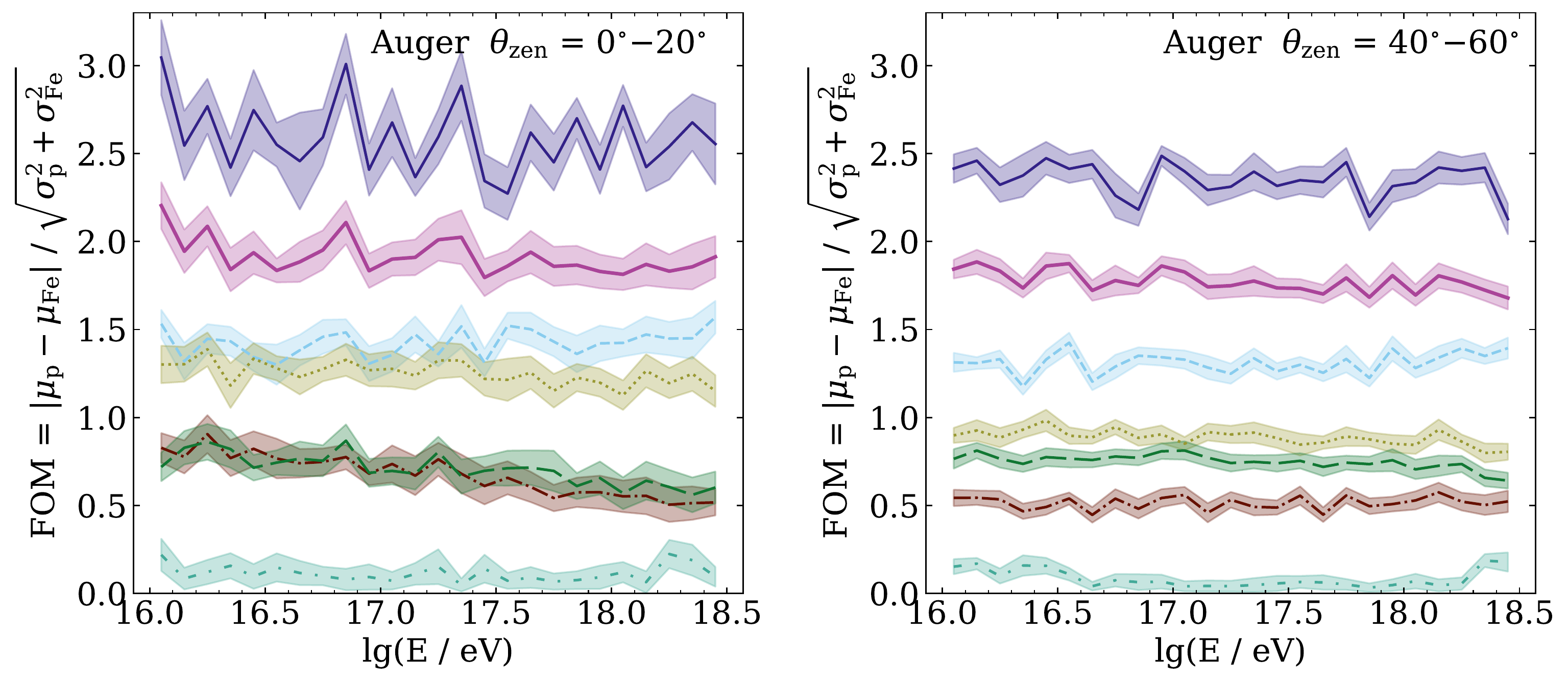}
\caption{Proton-iron separation power plots for all studied observables in this analysis at the sites of both the IceCube and Auger observatories. Included on each plot are the separation power curves showing what is possible when combined knowledge of all observables is known, either exactly or within assumed reconstruction uncertainties. All observables are scaled with respect to the energy reference observable prior to calculating the FOM curves.}
\label{fig:FOMs_combined_knowledge}
\end{figure*}

\section{Event-by-Event Mass Separation from Combined Knowledge of Air-Shower Observables} \label{sec:observables_combined}

Naturally, the overall separation power between primaries will increase with the addition of more mass sensitive observables.
Section~\ref{sec:sensitivity_w_fluctuations} used two-dimensional contour plots to exhibit the increase in separation power when adding knowledge of a second observable.
While higher-dimensional contour plots are difficult to visualize, the Fisher linear discriminant can easily be used to calculate the FOM for the combination of any number of shower observables.
As a quantitative estimate of the event-by-event mass separation between proton and iron primaries, the FOM was calculated as a function of binned primary particle energy for knowledge of each observable individually and when knowledge of all observables is combined.
The individual knowledge FOM values for all observables were calculated with knowledge of both the observable in question and the energy reference, $N_{\text{e, max}}$.

Fig.~\ref{fig:FOMs_combined_knowledge} shows the proton-iron separation power curves for each observable individually, along with their combined knowledge curves, at both the IceCube and Auger observatory locations.
FOM curves of the combination of all observables are shown twice, once with the assumed reconstruction uncertainties for all observables (magenta), and once with exact knowledge of all observables (dark blue).
Knowledge of individual observables were calculated within the baseline reconstruction uncertainties stated in Section~\ref{sec:IC_individual_observables}.
The left plots of Fig.~\ref{fig:FOMs_combined_knowledge} show the proton-iron separation power of observables using a zenith range of $\theta_{\text{zen}}$ = 0$^{\circ}-$20$^{\circ}$, while the right plots represent a zenith range of $\theta_{\text{zen}}$ = 40$^{\circ}-$60$^{\circ}$.
All FOM values, along with their uncertainty bands, were calculated from a bootstrap method, as described in Section~\ref{sec:IC_individual_observables}.
Due to limited shower statistics per energy bin, the proton-iron FOM curves for slightly inclined showers show large variations in separation power and larger uncertainty bands than the FOM curves for inclined showers.

As expected, exact knowledge of all observables yields the best proton-iron separation for both observatory locations.
Knowledge of all observables within their assumed baseline reconstruction uncertainties yields promising event-by-event mass separation between proton and iron primaries, with FOM values of approximately 2.0 at both IceCube and Auger for the zenith range of $\theta_{\text{zen}}$ = 40$^{\circ}-$60$^{\circ}$ and slightly higher separation power for nearly vertical showers within a zenith range of $\theta_{\text{zen}}$ = 0$^{\circ}-$20$^{\circ}$.
This dependence of the separation power on zenith angle is driven by the particles detectable at the surface.
In particular, the electron-muon ratio observables (for both high and low energies) and the low-energy surface muon observable provide large contributions, as there appears to be little to no zenith dependence on the \xmax, high-energy muon number, $L$, or $R$ observables.
Doubling the assumed baseline uncertainties for all observables decreases the combined knowledge FOM values by approximately 0.5 for both zenith ranges at the IceCube and Auger locations, resulting in lesser yet still promising event-by-event mass separation.

\subsection{Highly Mass Sensitive Observable Combinations} \label{sec:combinations}

Within this multivariate analysis, the combined knowledge mass separation curves are heavily influenced by only a handful of the studied observables.
The addition of more mass sensitive air-shower observables will improve mass separation; however, the individual separation power curves do not add linearly, making it hard to disentangle which combinations of observables contribute most to the overall mass separation curves in Fig~\ref{fig:FOMs_combined_knowledge}.
To study this, FOM values for observable pairs were calculated using Eq.~\ref{eq:figure_of_merit} for an energy range of 10$^{17.0}-$10$^{17.1}$\,eV and zenith range of 40$^{\circ}-$60$^{\circ}$.
All possible combinations of observable pairs are studied.
Table~\ref{tab:observable_combinations} shows the resulting FOM values for the observable pairs at the IceCube site, where all observables were scaled with respect to the energy reference observable.
The diagonal of Table~\ref{tab:observable_combinations}, shown in boldface font, represents the FOM for knowledge of only the scaled observable in question.
For reference, combining knowledge of all studied observables at the IceCube site for this energy and zenith range results in a FOM value of 2.8.

The highest FOM value corresponds to the observable combination of the high-energy muon number with the low-energy electron-muon ratio.
Although, the FOM value for this observable combination is only fractionally better than the combination of electrons at ground and high-energy muons.
Another notable observable pair is the combination of \xmax and low-energy surface muon number, as this combination yields the largest FOM value for any observable combination when excluding the high-energy muons.
Similar FOM values for these observable combinations are obtained at the Auger site; however, the separation power of any observable combination including low-energy muons is systematically higher at Auger.
As an example, the observable combination of \xmax and low-energy muon number at Auger results in a FOM value of 2.1, whereas the combination of \xmax and $N_{\text{e}}$ results in a FOM value of 1.4.

\begin{table*}[t!]
    \caption{FOM values calculated using Eq.~\ref{eq:figure_of_merit} for all possible pair combinations of observables at the IceCube site. All FOM values were calculated within an energy range of 10$^{17.0}-$10$^{17.1}$\,eV and zenith range of 40$^{\circ}-$60$^{\circ}$. All observables are scaled with respect to the energy reference observable prior to calculating the FOM values. The diagonal FOM values, listed in boldface font, correspond to knowledge of only that scaled observable.} \label{tab:observable_combinations}
    \vspace*{.5\baselineskip}
    \centering
    \begin{ruledtabular}
    \begin{tabular}{ c c c c c c c c c }
         & \multirow{2}{*}{\centering \xmax} & \multirow{2}{*}{\centering $N_{\text{e}}$} & \multirow{2}{*}{\centering $N_{\mu}$ E\,$>$\,500\,GeV} & \multirow{2}{*}{\centering $R_{\text{e}/\mu}$ E\,$>$\,500\,GeV} & \multirow{2}{*}{\centering $N_{\mu}$ 800\,m} & \multirow{2}{*}{\centering $R_{\text{e}/\mu}$ 800\,m} & \multirow{2}{*}{\centering $L$} & \multirow{2}{*}{\centering $R$} \\
         & & & & & & & & \\
        \colrule
         & & & & & & & & \\[-1.0em]
        \xmax & \textbf{1.4} & 1.4 & 2.1 & 1.4 & 1.9 & 1.5 & 1.5 & 1.5 \\
        $N_{\text{e}}$ & & \textbf{0.4} & 2.3 & 2.3 & 1.5 & 1.4 & 0.4 & 1.3 \\
        $N_{\mu}$ E\,$>$\,500\,GeV & & & \textbf{2.1} & 2.3 & 2.2 & 2.4 & 2.1 & 2.1 \\
        $R_{\text{e}/\mu}$ E\,$>$\,500\,GeV & & & & \textbf{0.9} & 1.7 & 1.0 & 0.9 & 1.4 \\
        $N_{\mu}$ 800\,m & & & & & \textbf{1.1} & 1.6 & 1.2 & 1.6 \\
        $R_{\text{e}/\mu}$ 800\,m & & & & & & \textbf{1.0} & 1.0 & 1.4 \\
        $L$ & & & & & & & \textbf{0.1} & 1.3 \\
        $R$ & & & & & & & & \textbf{1.3} \\
    \end{tabular}
    \end{ruledtabular}
\end{table*}


\section{Separation of Helium Against Proton and Oxygen Primaries} \label{sec:beyondProtonIron}

A multitude of cosmic-ray studies would benefit from the ability to discriminate between helium and the CNO group (carbon, nitrogen, oxygen), as well as proton and helium cosmic rays.
Measurements of the proton-to-helium ratio in the energy range of the Galactic to extragalactic transition, as well as the fraction of CNO, can help to distinguish certain scenarios for the origin of the highest energy Galactic cosmic rays~\cite{Thoudam:2016syr}.
Moreover, it is possible that these highest energy Galactic cosmic rays are rich in CNO while the extragalactic cosmic rays in the transition region may be rich in protons and possibly helium~\cite{PierreAuger:2022atd}.
In such scenarios, discrimination between cosmic rays of these mass groups (p, He, CNO) on an event-by-event basis may effectively enable to separate Galactic from extragalactic cosmic rays, allow for mass-group specific anisotropy measurements, and potentially enable rigidity-based particle astronomy of the most energetic Galactic sources.

The FOM metric was applied to measure the helium-oxygen and helium-proton separation power when using knowledge of the \xmax, $N_{\text{e, max}}$, $N_{\mu}$, $N_{\text{e}}$, $L$, and $R$ observables.
Similar to previous sections, the $N_{\mu}$ and $N_{\text{e}}$ observables were used to obtain the $R_{\text{e}/\mu}$ observable.
The use of observables within the 800$-$850\,m annulus allows the helium-oxygen and helium-proton separation powers to be obtained for both the IceCube Neutrino Observatory and Pierre Auger Observatory air-shower simulations.
For this section, high-energy muon observables will not be included on the plots as they lead to overcrowding, but the additional knowledge from these observables will still be discussed.

For this section, the FOM calculation becomes

\begin{figure*}[!t]
\centering
\includegraphics[trim={0in 0in 0in 0in}, clip, width=0.99\textwidth]{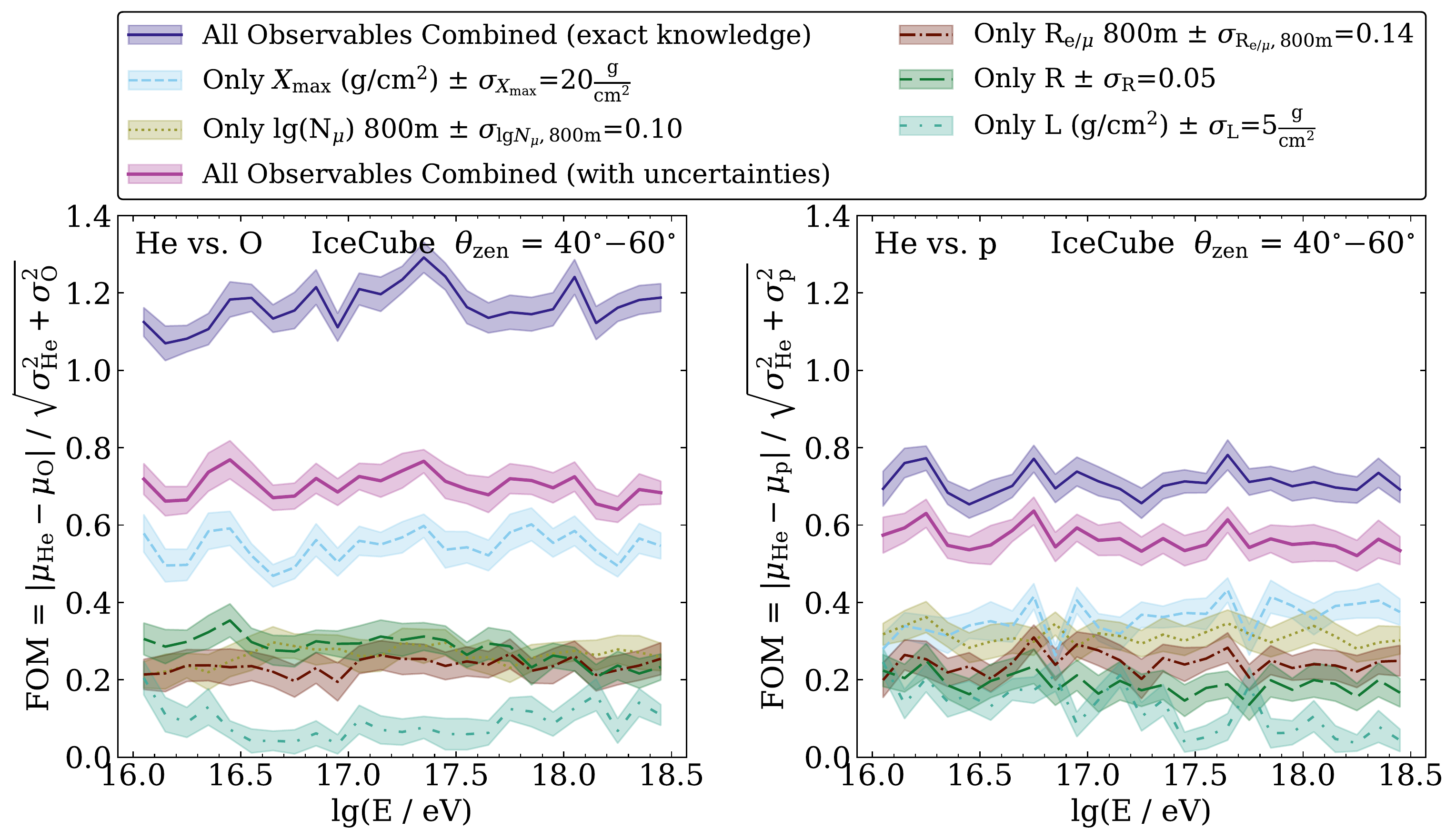}
\caption{Figure of merit plots showing the separation power of helium and oxygen primaries (left) and helium and proton primaries (right) at the IceCube Neutrino Observatory. Knowledge of high-energy muons is excluded to reduce crowding of lines on the plot. All observables are scaled with respect to the energy reference observable prior to calculating the FOM curves.}
\label{fig:beyondProtIron_FOM_plot}
\end{figure*}

\vspace{-10pt}

\begin{equation}
    \text{FOM} = \frac{| \mu_{\text{He}} - \mu_{\text{x}} |}{\sqrt{\sigma_{\text{He}}^{2} + \sigma_{\text{x}}^{2}}}. \label{eq:figure_of_merit_beyondProtIron}
\end{equation}

\noindent
where $\text{x}$ is replaced with p or O when calculating the separation power between helium and proton or helium and oxygen primaries respectively.
As before, the $\mu$ and $\sigma$ values represent the means and standard deviations of the primary particle distributions.
Fig.~\ref{fig:beyondProtIron_FOM_plot} presents these FOMs as a function of air-shower energy for the IceCube site, where the left plot shows the helium-oxygen separation power curves and the right plot shows the helium-proton curves.
The zenith angle range was chosen as $\theta_{\text{zen}}$ = 40$^{\circ}-$60$^{\circ}$ to maintain consistency with Fig.~\ref{fig:FOMs_combined_knowledge}.

For simulated air showers at IceCube, the helium-oxygen separation power is approximately two times lower than the proton-iron separation power for all FOM curves, with all curves showing similar relationships as a function of air-shower energy.
A similar result is observed with the cosmic-ray air showers from the Auger simulation library.
Although, Auger exhibits slightly larger helium-oxygen separation when knowledge of all observables is combined due to the increased separation power of the low-energy muons.
The additional knowledge of high-energy muons at IceCube raises the combined knowledge FOM curves to be similar to those of Auger; however, reasonable event-by-event helium-oxygen separation is not feasible at either observatory location.
The helium-proton separation power for both the IceCube and Auger observatories is systematically about three times lower than the proton-iron separation power for all plotted FOM curves, and therefore is less than the helium-oxygen separation power as well.
The addition of high-energy muon observables for the IceCube site adds less than 0.1 to the combined knowledge FOM curves for helium-proton separation.

One result from the helium-proton separation power curves in Fig.~\ref{fig:beyondProtIron_FOM_plot} is the $L$ curve does not contribute much to the overall helium-proton separation.
Hence, our results show $L$ is not a good observable for use in event-by-event mass discrimination between helium and proton primaries, although it may still be a good observable to determine the proton-helium ratio statistically for a large sample of events. 

The difference between $\ln(A)$ for helium and oxygen primaries is similar to this difference for helium and proton primaries, therefore the difference in $\mu$ from Eq.~\ref{eq:figure_of_merit_beyondProtIron} for these primary combinations is similar.
However, oxygen has smaller shower-to-shower fluctuations, leading to narrower distributions for all observables than for proton.
This smaller spread increases the separation between the helium and oxygen distributions, resulting in larger FOM values for the helium-oxygen separation power than those of helium-proton separation.


\section{Moving Beyond a Simple Fisher Analysis} \label{sec:GBDT}

Section~\ref{sec:beyondProtonIron} clearly illustrates the difficulty of separation for intermediate and lighter mass primaries on a per-event basis.
However, the Fisher analysis presented here only accounts for linear correlations between observables in each shower, and therefore a more advanced analysis technique could enhance the event-by-event mass separation beyond the one presented above.
A pipeline was established to train a gradient boosted decision tree (GBDT) for the regression task of predicting population zero-like and one-like events, where regression values of 0.0 and 1.0 represent populations zero and one respectively, corresponding to two different primary masses.
Several GBDTs were trained, where each individual model is trained and validated on data from a single combination of the three options: observatory location (IceCube or Auger), primary mass group separation (proton-iron, helium-oxygen, or helium-proton), and assumed observable reconstruction accuracy (exact knowledge or within assumed baseline uncertainties from Section~\ref{sec:IC_individual_observables}).
Each GBDT was defined using the scikit-learn python package~\cite{scikit-learn} where no hyperparameter optimization was performed.
Only default hyperparameters were used, other than the random seed used for each regression tree to control reproducibility.
Knowledge of all studied observables, \xmax, $R$, $L$, $N_{\text{e}}$, $N_{\mu}$, and $R_{\text{e}/\mu}$ (both high- and low-energy muon observables for the IceCube location only) were used as input features to the regression task.
Observables were scaled with respect to the energy reference observable, $N_{\text{e, max}}$, prior to training.
The same quality cuts for \xmax, $R$, and $L$ described at the end of Section~\ref{sec:simulations}, along with a zenith range from 40$^{\circ}-$60$^{\circ}$, were employed to ensure the same data was used in both the GBDT and Fisher analyses.

\begin{figure*}[!t]
\centering
\includegraphics[trim={0in 0in 0in 0in}, clip, width=0.99\textwidth]{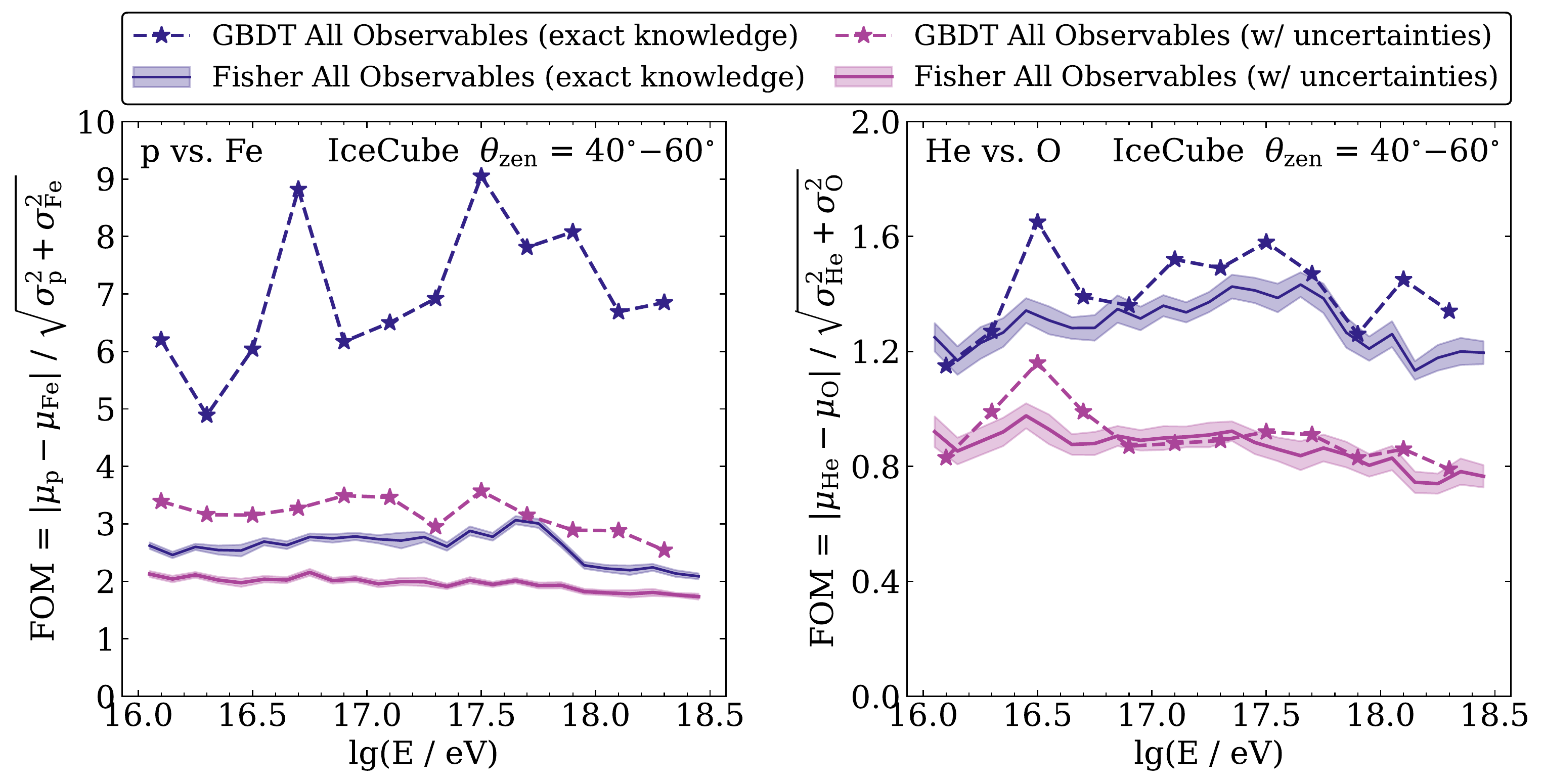}
\caption{Comparison of the Fisher and gradient boosted decision tree (GBDT) figure of merit values as a function of primary energy when knowledge of all observables is known either exactly (dark blue) or within the baseline reconstruction uncertainties assumed in this analysis (magenta). The left plot shows proton-iron separation while the right plot shows helium-oxygen separation, both at the location of the IceCube Neutrino Observatory. All observables are scaled with respect to the energy reference observable prior to calculating the FOM curves.}
\label{fig:GBDT_comparison}
\end{figure*}

A FOM was calculated using the validation output from the GBDTs following

\vspace{-10pt}

\begin{equation}
    \text{FOM} = \frac{| \mu_{\text{0}} - \mu_{\text{1}} |}{\sqrt{\sigma_{\text{0}}^{2} + \sigma_{\text{1}}^{2}}}. \label{eq:figure_of_merit_gbdt}
\end{equation}

\noindent
A training-validation ratio of 4:1 was used, where the validation set was further divided into primary energy bins of width $\log(E / \text{eV})$ = 0.2.
Using a wider energy binning than the Fisher analysis is necessary to ensure reasonable event statistics per energy bin.
Fig.~\ref{fig:GBDT_comparison} shows the FOM as a function of primary particle energy for the GBDT output using combined knowledge of all observables, both known exactly and within the assumed baseline reconstruction uncertainties, for the IceCube location.
Proton-iron separation is shown in the left plot of Fig.~\ref{fig:GBDT_comparison} while helium-oxygen separation is shown in the right plot.
The FOM values from the Fisher analysis are included for comparison.
For proton-iron separation, the GBDT performs substantially better than the Fisher analysis in the complete energy range studied within this analysis.
The GBDT with reconstruction uncertainties provides still greater proton-iron mass separation than the Fisher analysis with exact knowledge of all observables.
However, in terms of helium-oxygen separation, the GBDT shows little to no improvement over the simple Fisher analysis, with similar results observed for the helium-proton separation, and hence further highlights the difficulty of separating intermediate and light mass groups.
GBDTs trained and validated on simulations at the Auger location are not displayed here, as they show similar results.

For each trained GBDT, the observable importance was calculated from the purity gained by splitting the observable (i.e. the Gini importance).
For all GBDTs trained on simulations at the IceCube location, the observable yielding the most purity gain was the high-energy $N_{\mu}$ observable, followed immediately by either \xmax or the low-energy $N_{\mu}$ observable depending on the mass of the primary particles studied and the assumed observable reconstruction uncertainties.
For the Auger location, which excludes high-energy muons, the Gini importance determines the two most important observables to be \xmax and the low-energy muon number, where \xmax is most important when observable uncertainties are included and for exact knowledge proton-iron separation, yet low-energy $N_{\mu}$ is most important for the helium-oxygen and helium-proton separations when exact knowledge of the observables is known.
Table~\ref{tab:observable_gini_importance} lists the Gini importance of observables for proton-iron separation at the IceCube and Auger locations, both when observables are known exactly or known within reconstruction uncertainties.
Hence, the GBDT analysis performs equally well (helium-oxygen and helium-proton) to better (proton-iron) than the Fisher analysis, with consistencies for the most mass sensitive observables between the analysis methods.

\begin{table*}[t!]
    \caption{Gini importance of observables when separating proton and iron primaries with a GBDT at both the IceCube and Auger locations. Results when observables are known exactly (exact) or within reconstruction uncertainties (reco.) are shown. Importance values are normalized such that their sum is one within a column, where dashes (--) represent importance values of less than 0.005. The two observables with the most importance in each column are shown in bold. The importance values of high-energy muon observables at the Auger location are not available (N/A) as these observables were not studied at this location.} \label{tab:observable_gini_importance}
    \vspace*{.5\baselineskip}
    \centering
    \begin{ruledtabular}
    \begin{tabular}{ c c c c c }
         & \multirow{2}{*}{\centering IceCube p-Fe (exact)} & \multirow{2}{*}{\centering IceCube p-Fe (reco.)} & \multirow{2}{*}{\centering Auger p-Fe (exact)} & \multirow{2}{*}{\centering Auger p-Fe (reco.)} \\
         & & & & \\
        \colrule
         & & & & \\[-1.0em]
        \xmax & 0.01 & \textbf{0.09} & \textbf{0.67} & \textbf{0.79} \\
        $N_{\text{e}}$ & 0.01 & 0.01 & 0.01 & -- \\
        $N_{\mu}$ E\,$>$\,500\,GeV & \textbf{0.94} & \textbf{0.85} & N/A & N/A \\
        $R_{\text{e}/\mu}$ E\,$>$\,500\,GeV & -- & -- & N/A & N/A \\
        $N_{\mu}$ 800\,m & \textbf{0.04} & 0.04 & \textbf{0.25} & \textbf{0.16} \\
        $R_{\text{e}/\mu}$ 800\,m & -- & -- & 0.05 & 0.02 \\
        $L$ & -- & 0.01 & -- & 0.01 \\
        $R$ & -- & -- & 0.02 & 0.02 \\
    \end{tabular}
    \end{ruledtabular}
\end{table*}


\section{Discussion} \label{sec:discussion}

For the IceCube site, the high-energy muon observable provides the best mass separation for an \emph{individual observable}; however, for the highest energy air showers studied in this analysis, the IceCube \xmax observable shows similar or greater separation power depending on the assumed uncertainty of the observables.
The decrease in mass sensitivity of the high-energy muon observable above an energy threshold of 10$^{17.7}$\,eV was also observed in Fig.~\ref{fig:IC_observable_scaling}, as the means of the proton and iron distributions for this observable converge towards higher energies.
Potentially, at energies greater than 10$^{17.7}$\,eV, the relative energy fraction of the primary energy to the muon energy becomes relevant and allows for 500\,GeV muons to be produced at younger generations within the air shower.
This reduction in separation power of the high-energy muon observable should be studied in a future analysis.

While several experiments used or plan to use an electron-muon ratio for mass separation, we find that the FOM is lower than for the muon observables alone.
As all observables, including the muon numbers, have been scaled with the size of the electromagnetic component at the shower maximum, $N_{\text{e, max}}$, we conclude that there is no additional benefit in dividing the muon number by the electron number at ground $N_{\text{e}}$.
This implies experiments which can access $N_{\text{e, max}}$ or a similar observable via fluorescence, air-Cherenkov, or radio detection, may not have an immediate benefit from measuring the electron-muon ratio at ground.
Combining the electron and muon numbers in a Fisher linear discriminator shows at least a small improvement for the mass separation compared to the muon numbers alone, which indicates there are better ways of combining surface electron and muon observables than the simple ratio.

Regarding the \emph{combination of observables}, the simultaneous measurement of the muonic component and of \xmax enhances the per-event mass sensitivity.
At the IceCube site, the observable combination of high-energy muons and \xmax contributes most to the proton-iron separation power within most energy bins.
Replacing the high-energy muon observables with the low-energy surface muon observables within the 800$-$850\,m annulus results in lower, yet still promising, proton-iron separation power (at both the IceCube and Auger sites).
Nonetheless, for air-showers of EeV energies or above, the difference between using low- or high-energy muons becomes smaller, as the mass sensitivity of the high-energy muons decreases with energy.
Using \xmax in combination with both high- and low-energy muon observables increases the proton-iron separation power, but only fractionally when compared to having knowledge of only \xmax and high-energy muons.
Currently, IceCube can only reconstruct the low-energy surface muon observable when averaged over many events~\cite{IceCubeCollaboration:2022tla}, but with future analysis techniques and detection methods, per-event reconstructions may become feasible in IceCube-Gen2.

Within this analysis, the previously discussed observables were used to study the per-event mass separation, which is important, e.g., to search for mass-dependent anisotropies and to eventually enable particle astronomy through a per-event rigidity estimate.
A different, yet also important goal of air-shower experiments is to determine the average mass composition of incoming primary cosmic rays.
For the mass composition, averaging over large samples of events can reduce statistical uncertainties.
Thus, the per-event mass sensitivity becomes less important, as a lower mass sensitivity can be compensated by larger statistics.
Therefore, for measuring the mass composition, systematic uncertainties play a crucial role.
This includes uncertainties in the interpretation of air-shower observables when determining the absolute values of the atomic mass numbers.
The muon puzzle, a mismatch between air-shower simulations and muon measurements observed in several experiments~\cite{EAS-MSU:2019kmv, Soldin:2021wyv}, is an obvious indication that such systematic uncertainties exist at least regarding surface muon observables.
Hence, although muon observables will provide a higher event-by-event mass separation, experiments aiming at a low systematic uncertainty for the average mass composition should include \xmax sensitive detectors.

The lack of improvement from the gradient boosted decision tree over the Fisher analysis for intermediate and light mass separation highlights the difficulty of per-event mass sensitivity going forward.
Future studies need to show whether this is an intrinsic limitation or whether this can be overcome by more sophisticated, optimized GBDTs or by more advanced machine learning models, such as neural networks.
The latter could be better suited for this task, yet these methods pose their own risks, given the mismatch of different high-energy hadronic interaction models available today.


\section{Conclusion} \label{sec:conclusion}

Knowledge of multiple mass sensitive air-shower observables was obtained from CORSIKA simulations to study the event-by-event separation power of proton, helium, oxygen, and iron primary cosmic rays.
Observables studied include \xmax, the size of the electromagnetic shower component at \xmax and at ground level, and the muonic shower component at ground.
In addition, the potential mass sensitivity of the $L$ and $R$ shape parameters, from a parameterized Gaisser-Hillas fit to the longitudinal shower profile, was studied.
The choice of observables was motivated by detection methods of both current and planned air-shower arrays and was studied explicitly for air-shower simulations at locations of both the IceCube Neutrino Observatory and the Pierre Auger Observatory.

Promising event-by-event discrimination (FOM $>$ 1.5) between proton and iron primaries is seen at IceCube and Auger regardless of the energy or zenith angle ranges investigated in this analysis, even when including uncertainties.
This result holds for both a simple Fisher analysis and a gradient boosted decision tree, where the latter yields almost fully separable proton and iron distributions when exact knowledge of all observables is known.
At IceCube, the individual observable which yields the best proton-iron mass separation is the high-energy ($>$ 500\,GeV) muon number, measurable by the deep in-ice detector array; however, muons at such high energies cannot be distinguished in the underground muon detectors at Auger from muons of lower energy.
Overall, this analysis strengthens the science case for combining muon and \xmax detection at cosmic-ray observatories, as this observable combination proved to be highly mass sensitive.
The $R$ observable was also shown to be highly mass sensitive, especially when combined with muon or \xmax detection methods.
Furthermore, we confirm the special role of protons compared to nuclei for the $L$ observable~\cite{Buitink:2021pkz}.
Measuring $L$ with high precision and high statistics could be an interesting option to determine the proton-helium fraction in the mass composition, but we have shown that $L$ is not suited for event-by-event mass separation within the studied energy range.

Intermediate mass separation, such as helium-oxygen and helium-proton, is more difficult.
Both the Fisher linear discriminant and gradient boosted tree analyses presented in this work yield mediocre mass-separation powers for intermediate masses, even when combining all investigated shower observables and when assuming negligible uncertainties.
Hyperparamter optimization could improve the mass separation of the gradient boosted tree analysis, yet is prone to overfitting the training data.
Also, including finer details of the air showers, such as the azimuthal symmetry and smoothness of the radio and of the particle footprints, might further increase the mass separation.
These finer details, while important, require knowledge of air-shower array spacing, size, and sensitivity on a station level to determine reasonable reconstruction uncertainties.  
Hence, these observables were not studied here to provide an analysis of a more general purpose.

In summary, event-by-event separation of protons and iron nuclei as primary particles seems to be easily feasible with state-of-the-art techniques for air-shower arrays combining muon and \xmax detectors.
However, it is not clear from our study how well intermediate mass groups can be separated on an event-by-event basis because improving experimental resolutions on observables like \xmax or the muon number are insufficient due to the intrinsic shower-to-shower fluctuations.
Hence, an efficient event-by-event mass separation (if intrinsically feasible) necessarily requires the development of better analysis techniques.
That general conclusion holds over the complete range of studied energies from $10\,$PeV to over $3\,$EeV and the studied zenith angle range from $0^\circ$ to $60^\circ$ for both the Auger and IceCube sites.
There are smaller quantitative differences between both sites in the figures of merit, suggesting that the same air-shower array instrumentation could yield minute changes in the per-event mass separation based on location specific environmental factors (detector height above sea level, local geomagnetic field, atmospheric model, etc.).
Yet, we expect that our study can serve as a general guideline for the planning of future cosmic-ray arrays in the context of which detector combinations and reconstruction uncertainties are required to achieve mass separation on a per-event basis.
Although more detailed studies are required for array optimization at a particular site.


\begin{acknowledgements}

This research was supported in part through the use of Data Science Institute (DSI) computational resources at the University of Delaware.
This research was supported in part through the use of DARWIN computing system: DARWIN – A Resource for Computational and Data-intensive Research at the University of Delaware and in the Delaware Region, Rudolf Eigenmann, Benjamin E. Bagozzi, Arthi Jayaraman, William Totten, and Cathy H. Wu, University of Delaware, 2021, URL: https://udspace.udel.edu/handle/19716/29071.

The authors are very grateful to the Prague Auger group for providing the simulations for the paper.
The production of the simulations would not be possible without the use of the computing resources and the great support of the staff of the Computing Center of the Institute of Physics (CC IoP) of the Czech Academy of Sciences and of the DIRAC project.

We thank Stijn Buitink, Tim Huege, and Tanguy Pierog for useful discussions about this work, and we thank Anatoli Fedynitch and Roger Clay for useful suggestions improving the text.

Funding for this research was provided by the United States National Science Foundation through NSF award \#2046386.

Parts of this analysis were previously published in Ref.~\cite{Flaggs:2022} as a Master's thesis.

\end{acknowledgements}


\appendix

\section{Fisher analysis results for additional simulations at Auger} \label{appen:additional_auger}

As discussed in Section~\ref{sec:simulations}, Auger simulations beyond the energy range studied and with further high-energy hadronic interaction models are available and described in~\cite{PierreAuger:2021jov}.
Specifically, Sibyll 2.3c simulations extend up to energies of $10^{20.2}$\,eV while simulations were available from $10^{16.0}-10^{19.0}$\,eV for both the EPOS-LHC~\cite{Pierog:2013ria} and QGSJETII-04~\cite{Ostapchenko:2010vb} high-energy hadronic interaction models.
The Fisher analysis was repeated for the extended energy range of Sibyll 2.3c and the additional hadronic interaction models.
Observable scaling corrections with respect to the energy reference, $N_{\text{e, max}}$, have been recalculated separately for each hadronic model due to mismatches for observable predictions between models.
The observable scaling corrections for Sibyll 2.3c simulations have also been recalculated due to their larger energy range.

\begin{figure*}[!t]
\centering
\includegraphics[trim={0in 0in 0in 0in}, clip, width=0.99\textwidth]{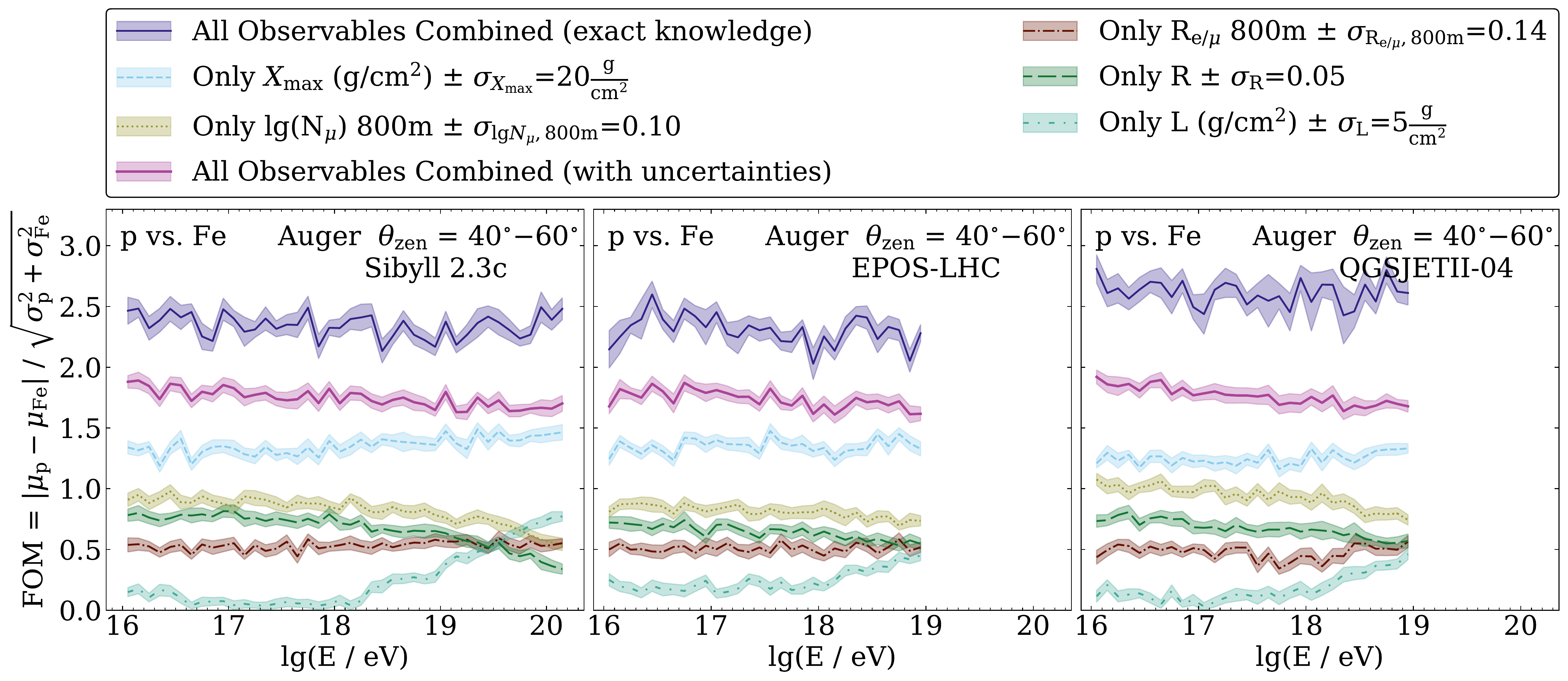}
\caption{Proton-iron figure of merit plots for Sibyll 2.3c (left), EPOS-LHC (middle), and QGSJETII-04 (right) simulations at the location of the Pierre Auger Observatory. The energy range of the Fisher analysis was extended corresponding to the simulations available for each hadronic model.}
\label{fig:hadronic_models}
\end{figure*}

Fig.~\ref{fig:hadronic_models} shows the proton-iron figure of merit curves calculated from the Fisher analysis using all studied observables for these additional Auger simulations.
A zenith angle range of $\theta_{\text{zen}}$ = 40$^{\circ}-$60$^{\circ}$ is shown, where uncertainties on individual observables remain the same as the baseline reconstruction uncertainties assumed in Section~\ref{sec:IC_individual_observables}.
Proton-iron mass separation curves are generally consistent between the three hadronic interaction models, both for individual observables and for combined knowledge of observables, except for the low-energy $N_{\mu}$ observable from the QGSJETII-04 model.
The low-energy muon observable for QGSJETII-04 exhibits a systematically higher proton-iron mass separation than both the Sibyll 2.3c and EPOS-LHC models, where this difference also increases the QGSJETII-04 mass separation curves when combined knowledge of all observables is known.

At energies above approximately $10^{18.5}$\,eV, all three hadronic interaction models exhibit a decrease in the proton-iron mass separation of both the low-energy $N_{\mu}$ and $R$ observables, while the mass separation of the $L$ observable increases.
Above $10^{19.7}$\,eV for the Sibyll 2.3c model, the proton-iron mass sensitivity of the $L$ observable becomes higher than the low-energy muon observable when reconstruction uncertainties are included.
However, when studying combinations of observables, the \xmax and low-energy $N_{\mu}$ combination remains the most mass sensitive.
Including the $L$ observable with both \xmax and low-energy $N_{\mu}$ increases the FOM by approximately 0.01, only.
Hence, while increasingly mass sensitive at the highest energies, $L$ adds little to no information about per-event mass separation if knowledge of both \xmax and surface muons is known.
These results for the additional Auger simulations hold for both the helium-oxygen and helium-proton separations, where helium-oxygen (helium-proton) FOMs are about a factor of two (three) less than proton-iron FOMs.
However, for helium-proton separation, $L$ as an individual observable does not exhibit an increase in separation power above $10^{18.5}$\,eV.

\section{Comparison between the use of parameterized Gaisser-Hillas and shifted Gaisser-Hillas fits to air-shower longitudinal profiles} \label{appen:fit_compare}

The $L$ and $R$ observables can also be derived from a typical Gaisser-Hillas function,

\vspace{-10pt}

\begin{equation}
    N = N_{\text{max}} \left( \frac{X - X_{0}}{X_{\text{max}} - X_{0}} \right)^{\frac{X_{\text{max}} - X_{0}}{\lambda}} \exp{\left( \frac{X_{\text{max}} - X}{\lambda} \right)}, \label{eq:GH_function}
\end{equation}

\noindent
using the definitions of $L$, $R$, and $X_{0}^{\prime}$ as defined in Section~\ref{sec:simulations}.
In Eq.~\ref{eq:GH_function}, $N$ represents the total number of electromagnetic particles at slant depth $X$.
$X_{0}$ and $\lambda$ are respectively related to the slant depth of first interaction and effective interaction length within the air shower, and thus hold no further physical significance.
Therefore, they are kept as free parameters in fits to air-shower longitudinal profiles.

Eq.~\ref{eq:GH_function} was also fit to all simulated air-shower longitudinal profiles, where the slant depth values ($X$) were shifted by $+$100\,$\text{g} \: \text{cm}^{-2}$.
The shift was performed to eliminate any potential negative $X - X_{0}$ values in the equation, which cause errors in the fits unless large portions of the longitudinal profile are removed from the simulations.
Shifting the slant depths had no impact on fit results when compared to a non-shifted Gaisser-Hillas fit, other than an increased total number of profiles for which fits can be obtained.
From the shifted fits, uncertainties in both $L$ and $R$ were found from error propagating their respective definitions and the same quality cuts of $\sigma_{\text{\xmax}} <$ 5\,$\text{g} \: \text{cm}^{-2}$, $\sigma_{L} <$ 5\,$\text{g} \: \text{cm}^{-2}$, and $\sigma_{R} <$ 0.05 were applied, as in Section~\ref{sec:simulations}.
Table~\ref{tab:fraction_poor_fits} shows the total percentage of fits which do not pass these quality cuts for both the shifted Gaisser-Hillas fit and the parameterized Gaisser-Hillas fit for all primaries simulated at the location of the IceCube Neutrino Observatory.

\begin{table}[!h]
    \caption{Percentage of total profile fits which do not pass the defined quality cuts for all simulated air showers at the location of the IceCube Neutrino Observatory. The shifted Gaisser-Hillas percentages, along with the parameterized Gaisser-Hillas percentages, are shown for each primary.} \label{tab:fraction_poor_fits}
    \vspace*{.5\baselineskip}
    \centering
    \begin{ruledtabular}
    \begin{tabular}{ c c c }
        Primary & GH Shift. & GH Param. \\
        \colrule
         & & \\[-1.0em]
        Proton & 19.6\% & 2.2\% \\
        Helium & 10.5\% & 0.9\% \\
        Oxygen & 3.6\% & 0.3\% \\
        Iron & 0.7\% & 0.1\% \\
    \end{tabular}
    \end{ruledtabular}
\end{table}

Both the shifted and parameterized Gaisser-Hillas fits to the air-shower longitudinal profiles yield similar values for the nominal fit parameters.
However, as illustrated in Table~\ref{tab:fraction_poor_fits}, the shifted Gaisser-Hillas fits are less stable.
Therefore the parameterized function was chosen as the nominal way to fit longitudinal profiles of air showers for this analysis.
The Auger simulation library shows similar results, although the total percentage of fits which fail the quality cuts is less for both fit types.
For the parameterized Gaisser-Hillas fits, there are slightly more iron showers which fail the quality cuts at Auger.
A possible explanation is the lack of CONEX in the air-shower simulations at Auger but this was not investigated in the analysis.
In addition to the quality cuts, $X_{0}$ and $\lambda$ are known to be highly correlated fit parameters~\cite{Andringa:2011zz}, making event-by-event analysis difficult and hence further motivating the use of a parameterized Gaisser-Hillas fit.

\section{Choice of annulus for the low-energy surface muon observables} \label{appen:annulus_FOM}

\begin{figure}[!th]
\centering
\includegraphics[trim={0in 0in 0in 0in}, clip, width=0.47\textwidth]{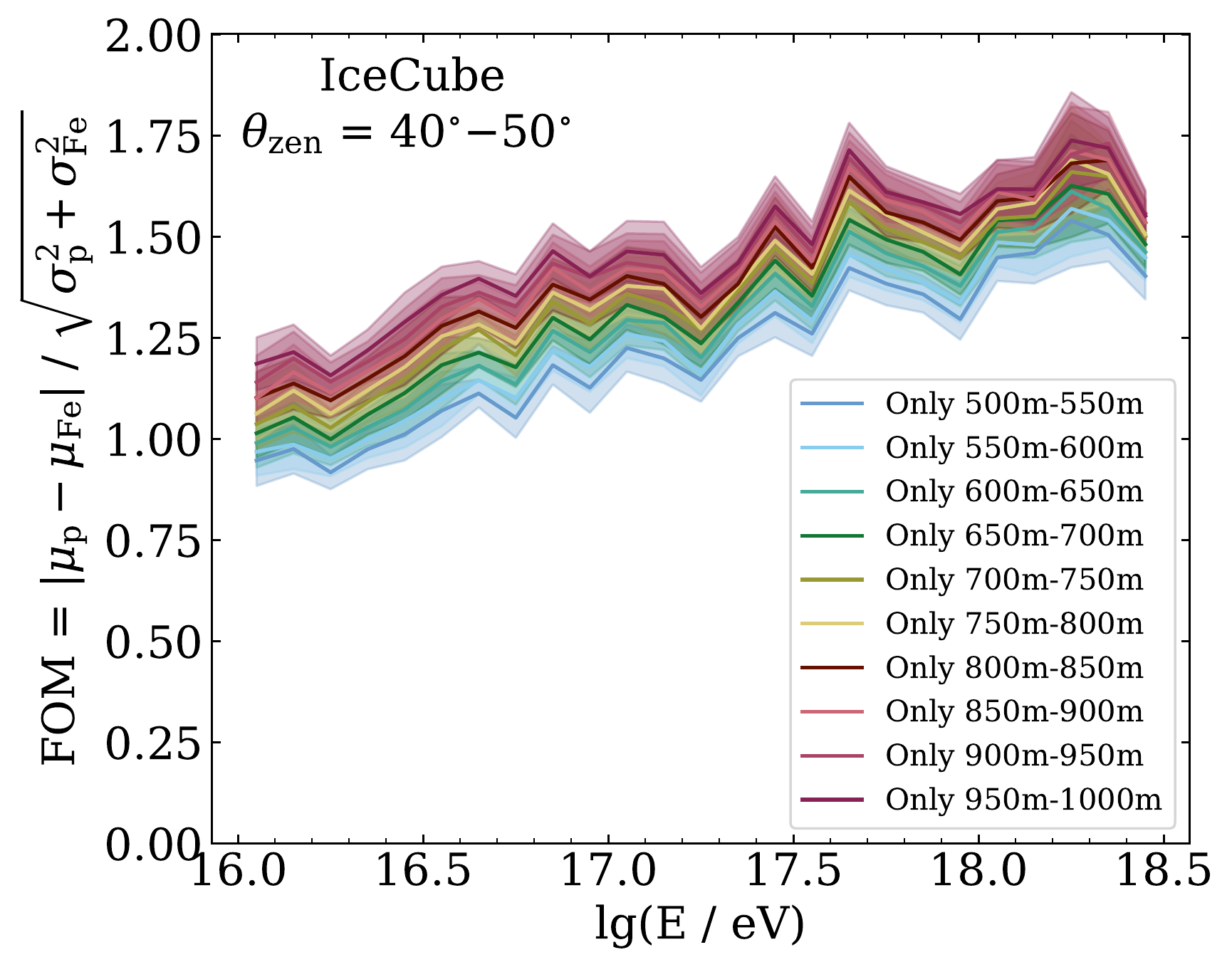}
\caption{Plot of proton-iron mass separation for IceCube simulated air showers using knowledge of only a low-energy muon observable within a 50\,m width annulus, placed at different distances to the air-shower axis. This FOM plot only shows the proton-iron separation power for the 10 annuli farthest from the air-shower axis.}
\label{fig:IC_muon_annuli_FOM}
\end{figure}

As discussed in Section~\ref{sec:scaling}, annuli of 50\,m width were investigated for the low-energy surface muon observables, with the farthest annulus ranging from 950$-$1000\,m from the air-shower axis.
A single annulus was determined for use in the mass sensitivity analysis by using the FOM metric to maximize proton-iron separation power while attempting to lessen the ratio of $N_{\mu}$ uncertainty-to-counts.
Due to the distribution of low-energy muons at ground and the Poissonian statistics of surface particle detectors then annuli closer to the shower axis will have smaller relative uncertainties.
The FOM curves as a function of air-shower energy for the muon number within the farthest 10 annuli from the shower axis are shown in Fig.~\ref{fig:IC_muon_annuli_FOM}.
Uncertainties in these FOM curves were estimated from a bootstrapping method, as described in Section~\ref{sec:IC_individual_observables}. 

The muon number within the farthest annulus exhibits maximum proton-iron separation power, although there are only small differences in separation power between this annulus and other annuli far from the shower axis.
Furthermore, based on IceTop observations, the muon densities at both 600\,m and 800\,m from the shower axis can be reconstructed at the South Pole~\cite{IceCubeCollaboration:2022tla}.
Motivated by all listed factors, the 800$-$850\,m annulus was chosen for use in the mass sensitivity analysis.
The FOM method was also applied to measure the proton-iron separation power of the low-energy $N_{\mu}$ observable for separate annuli at the Auger site.
The separation power curves follow a similar relationship as the curves for the IceCube simulation library shown in Fig.~\ref{fig:IC_muon_annuli_FOM}.
Therefore, to remain consistent with the analysis of IceCube simulations, the mass sensitivity of muon observables in the 800$-$850\,m annulus was investigated for Auger.

\section{Plots for air-shower observables at the Auger site} \label{appen:auger_plots}

As mentioned in the body of the paper, the same study of event-by-event mass separation performed at the site of IceCube was performed for air showers simulated at the Pierre Auger Observatory location.
A selection of the plots made for this study at the Auger site are included here for completeness.

\begin{figure}[!h]
\centering
\includegraphics[trim={0in 0in 0in 0in}, clip, width=0.47\textwidth]{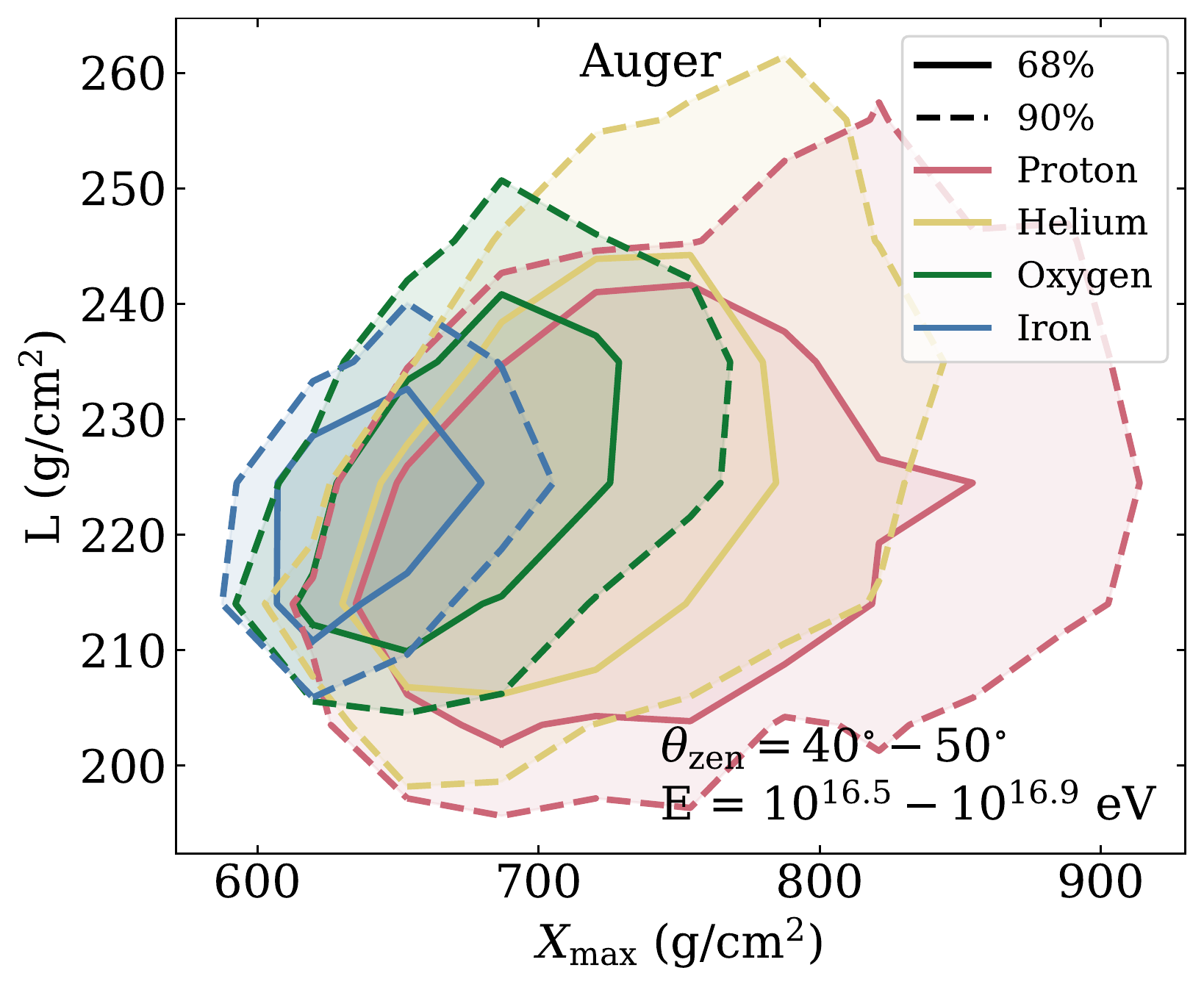}
\caption{Two-dimensional contour plot of the $L$ and \xmax air-shower observables from Pierre Auger Observatory simulations of all four primaries studied (proton, helium, oxygen, iron). Both the $L$ and \xmax observables are corrected with respect to the energy reference observable.}
\label{fig:auger_L_Xmax_contour}
\end{figure}

\begin{figure*}[!t]
\centering
\includegraphics[trim={0in 0in 0in 0in}, clip, width=0.808\textwidth]{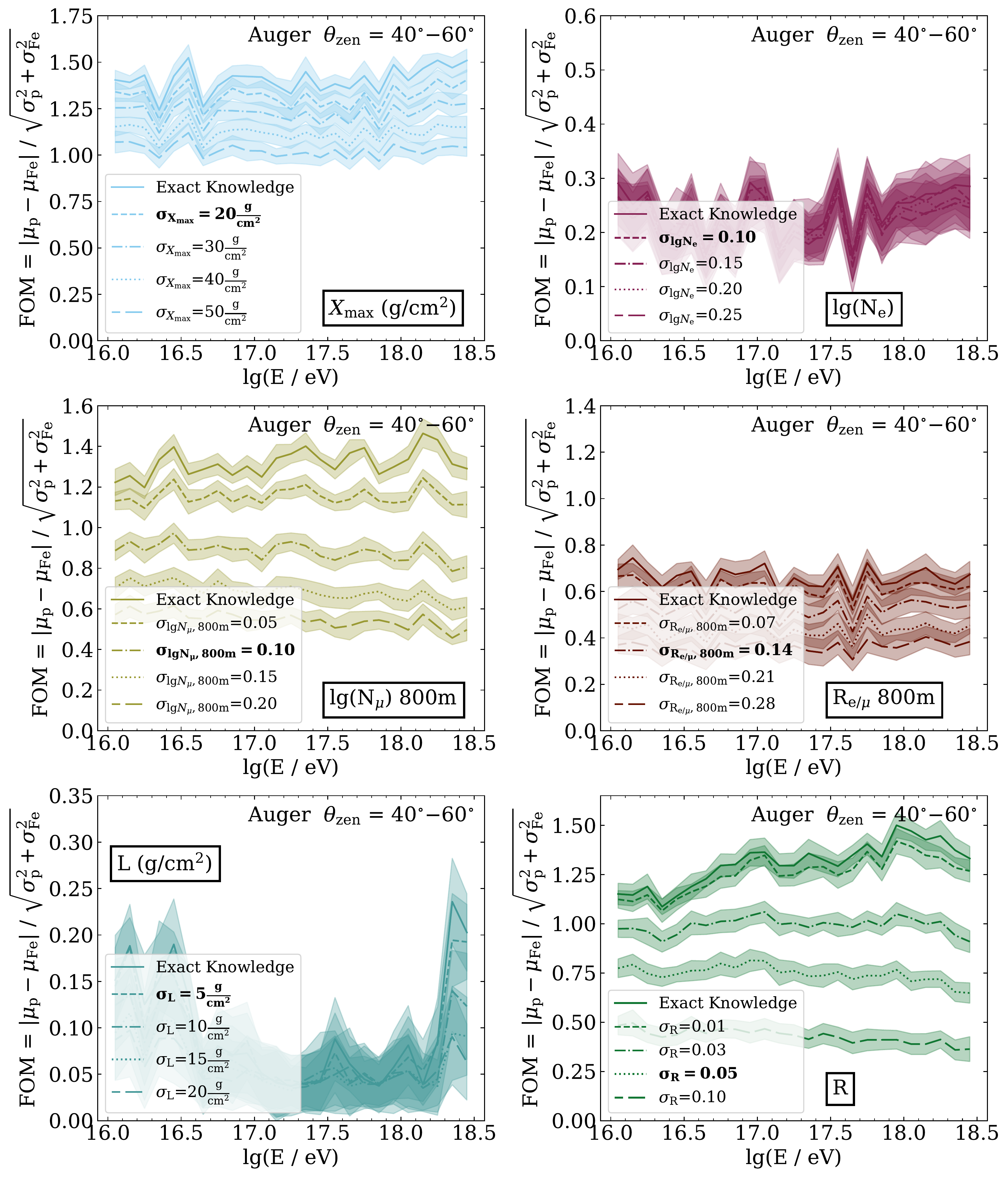}
\caption{Proton-iron separation power plots as a function of air-shower energy for knowledge of a single air-shower observable. A range of reconstruction uncertainties for each observable are assumed and their respective separation power curves are included within the plots, where baseline uncertainties are listed in boldface font. Only observables investigated at the site of Auger are included, therefore the high-energy muon observables are not included. Scalings of the y-axes are the same as in Fig.~\ref{fig:IC_single_observables} to allow direct comparison between the figures. All observables are scaled with respect to the energy reference observable.}
\label{fig:auger_single_observables}
\end{figure*}

\bibliographystyle{apsrev4-1}
\bibliography{references}

\end{document}